\documentclass[aps,twocolumn,amsmath,superscriptaddress,prd,floatfix]{revtex4-2}
\usepackage{graphicx,color,dcolumn,booktabs,bm}
\usepackage{txfonts}
\usepackage{overpic}
\usepackage{caption}
\usepackage{epsfig}
\usepackage{amssymb}
\usepackage{epstopdf}
\usepackage{indentfirst}
\usepackage{feynmf}
\usepackage{slashed}
\usepackage{cases}
\usepackage{color}
\usepackage{multirow}
\usepackage{makecell}
\usepackage{float}
\usepackage{graphicx,color,dcolumn,booktabs,bm}
\usepackage{cases}
\usepackage{array}

\graphicspath{{Figures/}} 

\usepackage[colorlinks,citecolor=blue,anchorcolor=red,menucolor=red,linkcolor=red,filecolor=red,runcolor=red,urlcolor=blue,frenchlinks=red]{hyperref}

\begin{document}

\title{Radiative decay of heavy-light mesons from lattice QCD}
\affiliation{Guangdong Provincial Key Laboratory of Nuclear Science, Institute of Quantum Matter, South China Normal University, Guangzhou 510006, China}

\author{Wen-Zheng Hou} %
\author{Nan Wang} %
\affiliation{State Key Laboratory of Nuclear Physics and Technology, Institute of Quantum Matter, South China Normal University, Guangzhou 510006, China}
\affiliation{Key Laboratory of Atomic and Subatomic Structure and Quantum Control~(MOE), Guangdong-Hong Kong Joint Laboratory of Quantum Matter, Guangzhou 510006, China}
\affiliation{Guangdong Basic Research Center of Excellence for Structure and Fundamental Interactions of Matter, Guangdong Provincial Key Laboratory of Nuclear Science, Guangzhou 510006, China}
\author{Long-cheng Gui} %
\email{guilongcheng@hunnu.edu.cn}
\affiliation{Department of Physics, Hunan Normal University,  Changsha 410081, China }
\affiliation{Synergetic Innovation
Center for Quantum Effects and Applications (SICQEA), Changsha 410081, China}
\affiliation{Key Laboratory of Low-Dimensional Quantum Structures and Quantum Control of Ministry of Education, Changsha 410081, China}
\author{Jun Hua} %
\author{Jian Liang} %
\email{jianliang@scnu.edu.cn}
\author{Jun Shi} %
\email{jun.shi@scnu.edu.cn}
\affiliation{State Key Laboratory of Nuclear Physics and Technology, Institute of Quantum Matter, South China Normal University, Guangzhou 510006, China}
\affiliation{Guangdong Basic Research Center of Excellence for Structure and Fundamental Interactions of Matter, Guangdong Provincial Key Laboratory of Nuclear Science, Guangzhou 510006, China}
\author{Yu Meng} %
\email{yu\_meng@zzu.edu.cn}
\affiliation{School of Physics, Zhengzhou University, Zhengzhou, Henan 450001, China\\
({CLQCD Collaboration})}

\begin{abstract}
We present the first systematic study of the radiative decays of charmed mesons using $2+1$-flavor clover fermion gauge ensembles generated by the CLQCD collaboration. One of the ensembles is at the physical pion mass, and one has a fine lattice spacing $a\sim 0.05 ~\text{fm}$. We determine the coupling constants to be $g_{D^{\ast+} D^+ \gamma} = -0.204(22)$ GeV$^{-1}$, $g_{D^{\ast0} D^0 \gamma} = 1.73(37)$ GeV$^{-1}$, and $g_{D_s^{\ast+} D_s^+ \gamma} =-0.120(14)$ GeV$^{-1}$, respectively.
Compared with previous studies, our results demonstrate significant improvements in precision. In particular, we carefully estimate the systematic uncertainty arising from matrix element fits, momentum transfer extrapolations, and chiral and continuum limit extrapolations, which are included in the reported total uncertainties. These couplings yield the following predictions of decay widths: $\Gamma_{D^{\ast+} \rightarrow D^+ \gamma} = 0.253(55)$ keV, $\Gamma_{D^{\ast0} \rightarrow D^0 \gamma} = 18.2(7.8)$ keV, and $\Gamma_{D_s^{\ast+}\rightarrow D_s^+ \gamma} = 0.094(22)$ keV. This work establishes
first-principles results of the charmed meson radiative transitions and provides inputs for understanding the structure and properties of heavy-light mesons.
\end{abstract}

\maketitle

\section{Introduction}

The transition form factors and coupling constants of radiative decays encode the internal electromagnetic structure of hadrons. However, current experimental studies of the radiative decays of the $D^{\ast}$ and $D_s^{\ast}$ mesons primarily focus on measuring the branching fractions, and only the $D^{\ast+}$ has a precisely measured total decay width. Therefore, for the rest of the $D^{\ast}$ and $D_s^{\ast}$ mesons, the coupling constants cannot be directly determined from experimental data.
For the branching fractions,
the most recent result of the $D^{\ast+}$ meson is $\mathcal{B}(D^{\ast+} \to D^+ \gamma) =(1.68\pm0.42\pm0.29)\%$, which was measured in 1998 by the CLEO collaboration~\cite{CLEO:1997rew}.
As for the neutral counterpart $D^{\ast0}$, BESIII obtained $\mathcal{B}(D^{\ast0} \to D^0 \gamma) = (34.5\pm0.8\pm0.5)\%$ in 2014~\cite{BESIII:2014rqs}. However, there is no direct measurement of the branching fraction for the $D_s^{\ast}$ meson yet; only a ratio of branching fractions $\mathcal{B}(D_s^{\ast+} \to D_s^+\pi^0)/\mathcal{B}(D_s^{\ast+} \to D_s^+ \gamma) = (6.16\pm0.43\pm0.18)\%$ is available from BESIII in 2022~\cite{BESIII:2022kbd}.

Theoretically, the radiative decays of heavy-light mesons have been studied using various approaches, such as QCD sum rules (QCDSR), light cone sum rules (LCSR), and various quark models~\cite{Li:2020rcg, Pullin:2021ebn, Amundson:1992yp, Tran:2023hrn, Cheung:2014cka, Orsland:1998de, Jaus:1996np, Goity:2000dk, Aliev:1994nq,Lu:2024tgy}. In parallel, the lattice QCD community has provided several first-principles results. For the $D^\ast$ meson, $g_{D^{\ast+}D^+\gamma}=-0.2(3)~\text{GeV}^{-1}$ and $g_{D^{\ast0}D^0\gamma}=2.0(6)~\text{GeV}^{-1}$ were reported based on calculations with a chiral extrapolation but at a fixed lattice spacing of $\beta=5.40$~\cite{Becirevic:2009xp}. For the $D^\ast_s$ meson, direct lattice computations yielded effective form factors $V_{\text{eff}}(0) = -0.20(4)$~\cite{Donald:2013sra} and $V_{\text{eff}}(0) = -0.178(9)$~\cite{Meng:2024gpd}, using which the corresponding coupling constant can be obtained by $g_{VP}=2V_{\text{eff}}(0)/(m_V+m_P)$. The former study~\cite{Donald:2013sra} performed both chiral and continuum extrapolations but did not include an ensemble at the physical pion mass. The latter one~\cite{Meng:2024gpd} carried out only the continuum extrapolation. Additionally, an indirect lattice determination from studying the $D^+_s \to l^+\gamma\nu_l$ process gives $g_{D_s^{\ast +} D^+_s \gamma} =-0.118(13)~\text{GeV}^{-1}$~\cite{Frezzotti:2023ygt}.

In this work, we calculate the form factors and the couplings of the radiative decays $D^{\ast+} \to D^+ \gamma$, $D^{\ast0} \to D^0 \gamma$, and $D_s^{\ast+} \to D_s^+ \gamma$. 
Our calculation utilizes 6 gauge ensembles of clover fermions generated by the CLQCD collaboration~\cite{CLQCD:2023sdb}, including one ensemble with a fine lattice spacing of 0.052 fm and one ensemble at the physical pion mass.
We perform simultaneous chiral and continuum extrapolations to obtain the final predictions.
The systematic uncertainties, involving the matrix element fit, the momentum transfer extrapolation, and the combined chiral and continuum limit extrapolation,
are all carefully estimated.

We determine the radiative decay couplings as $g_{D^{\ast+} D^+ \gamma} = -0.204(22)$ GeV$^{-1}$, $g_{D^{\ast0} D^0 \gamma} = 1.73(37)$ GeV$^{-1}$, and $g_{D_s^{\ast+} D_s^+ \gamma} =-0.120(14)$ GeV$^{-1}$. Based on these couplings,
we calculate the corresponding radiative decay widths. Using the experimentally measured total width $\Gamma(D^{\ast+})$ from Ref.~\cite{BaBar:2013thi} as an input, we obtain $\Gamma(D^{\ast+} \to D^+ \gamma) = 0.253(55)$ keV, corresponding to a branching fraction $\mathcal{B}(D^{\ast+} \to D^+ \gamma) = 0.30(7)\%$. This result is significantly lower than the CLEO 1998 measurement of $(1.68\pm0.42\pm0.29)\%$. The reason can be attributed to some unknown systematic uncertainties on either side, which can be explored through further studies. For the $D^{\ast0}$ case, we get $\Gamma(D^{\ast0} \to D^0 \gamma) = 18.2(7.8)$ keV. By combining this with the experimental branching fraction, we predict the total width to be $\Gamma_{D^{\ast0}} = 52(22)$ keV. Similarly, using the coupling $g_{D_s^{\ast+} D_s^+ \gamma}$, we derive $\Gamma(D_s^{\ast+} \to D_s^+ \gamma) = 0.094(22)$ keV. By taking the known branching ratio into account, our result infers that the total decay width is $\Gamma(D_s^{\ast+}) = 0.100(24)$ keV.

The rest of this article is organized as follows: In Sec.~\ref{sec2}, we detail the theoretical framework, including the definitions of relevant correlators, the formulae for extracting effective form factors from matrix elements, and the procedures for calculating the corresponding decay widths. The numerical results are presented in Sec.~\ref{sec3}. Specifically, Sec.~\ref{sec31} discusses the hadron spectrum and the dispersion relation. Sec.~\ref{sec32} describes the joint fit method for matrix element extraction and the estimation of the corresponding systematic uncertainties. In Sec.~\ref{sec33}, we perform the momentum transfer extrapolations using the $z$-expansion formula and also estimate the related systematic uncertainty. Sec.~\ref{sec34} discusses the combined chiral and continuum extrapolations. Finally, we summarize our results in Sec.~\ref{sec4} and provide supplementary plots and data in the Appendix.

\section{Formalism} \label{sec2}

The radiative transition matrix elements of processes $V\to P\gamma$ can be extracted from Euclidean correlation functions. 
The two-point functions with momentum $\vec{p}$ for vector mesons with Lorentz index $i$ and for pseudoscalar mesons are 
\begin{equation}
    C_{2i}^V(t) = \sum_{\vec{x}} e^{-i\vec{p}\cdot\vec{x}} \left\langle 0 \left| \mathcal{O}^V_i(\vec{x},t)\ \mathcal{O}^{V\dagger}_i(0) \right| 0 \right\rangle, 
\end{equation}
and
\begin{equation}
    C_{2}^P(t) = \sum_{\vec{x}} e^{-i\vec{p}\cdot\vec{x}} \left\langle 0 \left| \mathcal{O}^P(\vec{x},t)\ \mathcal{O}^{P\dagger}(0) \right| 0 \right\rangle,
\end{equation}
respectively, where $\mathcal{O}^V_i = \bar{l}\gamma_i c$ and $\mathcal{O}^P = \bar{l}\gamma_5 c$. The light quark field $l$ denotes \( u \), \( d \), or \( s \), and \( c \) represents the charm quark field. By inserting a current operator, one constructs a three-point correlation function:
\begin{equation}
C_{3\mu i}(t,\tau) = \sum_{\vec{x},\vec{y}} e^{-i\vec{p}\cdot\vec{x}-i\vec{q}\cdot\vec{y}} \left\langle 0 \left| \mathcal{O}^P(\vec{x},t)\ J_\mu(\vec{y},\tau)\ \mathcal{O}^{V\dagger}_i(0) \right| 0 \right\rangle,
\end{equation}
where $\tau$ denotes the current insertion time, $J_\mu$ is the current operator, and its explicit form is presented in Sec.~\ref{sec32}. After inserting complete sets of eigenstates into the correlation functions, we arrive at
\begin{equation}
 C^V_{2i}(t) =\sum_n \frac{1}{2E^V_n} \left|A^V_{n,i}\right|^2 \left( \frac{1+{p^V_i}^2}{{m^V_n}^2}\right) e^{-E^V_n t}, \label{eq3-1} 
\end{equation}
\begin{equation}
C_2^P(t) =\sum_m \frac{1}{2E^P_m} \left|A^P_{m}\right|^2 e^{-E^P_m t}, \label{eq3-2}
\end{equation}
and
\begin{equation}
C_{3\mu i}(t,\tau)
= \sum_{m,n}\frac{A_{m}^P A^{V\ast}_{n,i}\left\langle P_m \left| J_\mu \right| V_{n,i} \right\rangle  e^{-(E^V_{n}-E^{P}_m) \tau-E^{P}_m t}}{2E^V_n 2E^P_m}.
\label{eq3}
\end{equation}
This procedure
decomposes the correlation functions into a combination of matrix elements and kinematic factors. Specifically, $A^V_{n,i} = \left\langle 0 \left| O^V_i \right| V_{n,i} \right\rangle$ is the matrix element between the vacuum and the $n$-th vector meson state, and the matrix element $A^P_{m} = \left\langle 0 \left| O^P \right| P_{m} \right\rangle$ is defined analogously for the pseudoscalar meson states.
The factor $\frac{1 + {p_i^V}^2}{{m_n^V}^2}$ originates from the polarization sum $\sum_r \epsilon_i(r,p)\epsilon^\ast_j(r,p) =\delta_{ij}+\frac{p_ip_j}{m^2}$. The quantity  $\left\langle P_m \left| J_\mu \right| V_{n,i} \right\rangle$ represents the transition matrix element between the $m$-th pseudoscalar state and the $n$-th vector state. In the present study, we focus primarily on the ground-state matrix element $\left\langle P \left| J_\mu \right| V_{i} \right\rangle$, where the ground-state indices \( m = n = 0 \) are omitted for notational simplicity.
Taking into account the constraints imposed by Lorentz covariance and the Ward identity, the transition matrix elements can be parameterized as
\begin{equation}
    \left\langle P(p') \left| J_\mu(q) \right| V_{i}(p) \right\rangle = \frac{2V_{\text{eff}}(Q^2)}{m^{P} + m^{V}} \epsilon_{\mu i\alpha\beta} p'^\alpha p^\beta,
    \label{eq4}
\end{equation}
where $q = p - p'$, $Q^2=-q^2$, and $V_{\text{eff}}(Q^2)$ are the effective transition form factors that encode the dynamical information of this process. Along with the corresponding kinematic factors and phase space, the form factor can be used to calculate the radiative decay width. The transition amplitude for the radiative decay $V \to P \gamma$ is given by
\begin{equation}
\mathcal{M}_{\lambda\lambda^{\prime}} = e\  \epsilon_{V}^{\mu}(p,\lambda)\epsilon_{\gamma}^{i}(q,\lambda)\left\langle P(p^{\prime}) \left| J_{\mu}(q) \right| V_{i}(p) \right\rangle,
\end{equation}
where $\lambda^\prime$ and $\lambda$ denote the polarization indices of the vector meson and the photon, respectively.

After averaging over the initial vector meson's polarizations and summing over the final photon's polarizations, one obtains
\begin{equation}
\frac{1}{3} \sum_{\lambda, \lambda^{\prime}} \left|\mathcal{M}_{\lambda\lambda^{\prime}}\right|^{2} = \frac{e^{2}}{6} \left( \frac{2V_{\rm eff}(0)}{m_{V}+m_{P}} \right)^{2} \left(m_{V}^{2} - m_{P}^{2}\right)^{2},
\end{equation}
where the prefactor $1/3$ accounts for the polarization average of the initial vector meson. In the end, the coupling constant $g_{VP}$ of the radiative process is given by $g_{VP}={2V_{\text{eff}}(0)}/{\left(m_V+m_P\right)}$.
The decay width is then obtained by integrating the squared amplitude over the two-body phase space. In the rest frame of the vector meson, the width can be expressed as
\begin{equation}
\Gamma = \frac{m^2_V-m_P^2}{64\pi^2 m_V^3} 
\int \left( \frac{1}{3} \sum_{\lambda, \lambda^{\prime}} \left|\mathcal{M}_{\lambda\lambda^{\prime}}\right|^{2} \right) d\Omega.
\end{equation}
And the final result reads:
\begin{equation}
\Gamma = \frac{\alpha \left(m_{V}+m_{P}\right)\left(m_{V}-m_{P}\right)^{3} V_{\rm eff}^{2}(0)}{6 m_{V}^{3}},
\end{equation}
after the integration is explicitly carried out, where $\alpha = e^{2}/(4\pi)$ is the fine-structure constant.

\section{Numerical details} \label{sec3}

\begin{table*}[t]   
\centering
\begin{tabular}{lccccccccc}
\toprule
Ensemble & $a$ (\text{fm}) & $m_\pi$ (\text{MeV}) & $n_{\text{cfg}}$ & $am_s$ &$am_c$ & $L^3 \times T$ & $Z_V$ & $n_{\text{size}}$ & $n_{\text{iter}}$ \\
\midrule
F32P21 & 0.07746(18) & 210.9(2.2) & 456 & -0.2050& 0.2104 & $32^3 \times 64$ & 0.83579(09) & 3 & 120 \\
C32P23 & 0.10530(18) & 228.0(1.2) & 445 & -0.2400& 0.4500 & $32^3 \times 64$ & 0.79957(13) & 2 & 120 \\
H48P32 & 0.05187(26) & 317.2(0.9) & 905 & -0.1700& 0.0581 & $48^3 \times 144$ & 0.86855(04) & 4 & 120 \\
F32P30 & 0.07746(18) & 303.2(1.3) & 776 & -0.2050& 0.2079 & $32^3 \times 64$ & 0.83548(12) & 3 & 120 \\
C24P29 & 0.10530(18) & 292.7(1.2) & 948 & -0.2400& 0.4479 & $24^3 \times 72$ & 0.79810(13) & 2 & 120 \\
C48P14 & 0.10530(18) & 135.5(1.4) & 448 & -0.2310& 0.4502 & $48^3 \times 96$ & 0.79957(06) & 2 & 120 \\ 
\bottomrule
\end{tabular}
\caption{This table lists the parameters of the ensembles used in the calculation, including the lattice spacings $a$, number of configurations $n_{\text{cfg}}$, pion masses $m_{\pi}$, volumes $L^3 \times T$, quark masses, and the vector-current renormalization constants $Z_V$. The last two columns provide the parameters for Gaussian smearing. The quark sources are created at the $t=0$ time slice on each ensemble in the practical computation.}
\label{tab:lattice_params}
\end{table*}

This work utilizes $2+1$-flavor clover fermion gauge ensembles generated by the CLQCD collaboration~\cite{CLQCD:2023sdb}. The charm quark mass parameters are the same as those in Ref.~\cite{Meng:2024gpd}, which were determined by recovering the physical $J/\psi$ mass. The lattice spacings, volumes, and the vector-current renormalization constants $Z_\text{V}$ are detailed in Table~\ref{tab:lattice_params}.  

We use the sink-sequential scheme 
to generate three-point functions.
We apply Gaussian smearing at both the source and the sink when computing quark propagators,
which is necessary to suppress the excited-state
contamination and benefits the joint fit of the two-point and three-point functions.
The smearing parameters are also listed in Table~\ref{tab:lattice_params}. 
When using the sequential source technique, three different source-sink separations (denoted as $t$) were calculated on each ensemble. The three momenta of the final-state include two types: $\vec{p}'=(0,0,0)\left({2\pi}/{L}\right)$ and $\vec{p}'=(0,0,1)\left({2\pi}/{L}\right)$, and the maximum momentum transfer is set to $q^2= 3\left({2\pi}/{L}\right)^2$. The two-point functions incorporate momenta up to $p^2= 6\left({2\pi}/{L}\right)^2$.


\subsection{Masses and the dispersion relation}\label{sec31}

\begin{figure}[h]
    \centering
    \includegraphics[width=0.45\textwidth, keepaspectratio]{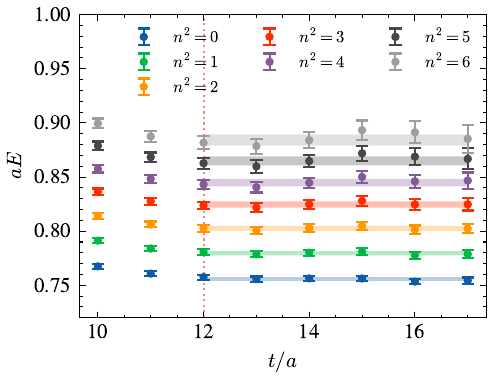}   
    \caption{The data points represent the effective energy extracted from two-point function on the F32P21 ensemble using Eq.~(\ref{corr}), and the bands correspond to the fitted result using Eq.~(\ref{single}). The $n^2$ in the legend labels the discrete momenta $p^2 = n^2\left({2\pi}/{L}\right)^2$. The dashed vertical line indicates the starting point of the fits.}\label{fig1}  
\end{figure}

The effective energy at each time slice can be extracted using
\begin{equation}
aE(t)=\ln{\frac{C_2(t)}{C_2(t+1)}}.
\label{corr}
\end{equation}
from the two-point functions. When a plateau emerges, it indicates that the two-point functions are dominated by the ground-state. We, therefore, fit the correlators with different momenta from the starting point of the plateau using a single-exponential form
\begin{equation}
C_2(t) = A_0 e^{-aE_0 t}.
\label{single}
\end{equation}
Fig.~\ref{fig1} shows the effective energies and the corresponding fit results of the $D$ meson on the F32P21 ensemble. We start the fits at time slice $t/a=12$, and the correlations between time slices are taken into consideration. The fit results are shown as bands, with uncertainties that are comparable to those of the chosen starting data points. The subsequent data points almost all lie within one sigma of the fitted bands.

\begin{figure}[h]
    \centering
    \includegraphics[width=0.45\textwidth, keepaspectratio]{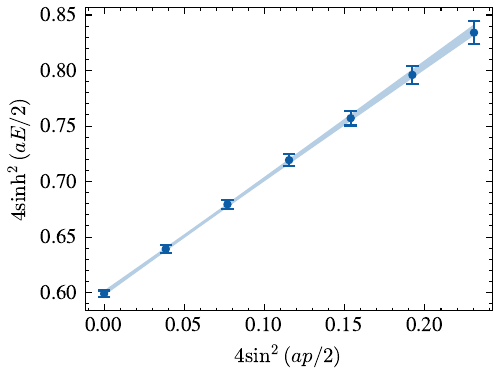}
    \caption{Dispersion relation results for the $D$ meson on the F32P21 ensemble. The data points are obtained under the fit scheme as shown in Fig.~\ref{fig1}. After converting the lattice momentum $ap$ to $2\sin(ap/2)$ and the lattice energy  $aE$ to $2\sinh(aE/2)$, the slope in the plot, which corresponds to $c^2$, is consistent with one within 2 sigmas.}\label{fig2}  
\end{figure}  

\begin{table*}[t]
\small
\centering
\setlength{\tabcolsep}{3pt} 
\begin{tabular}{@{}lcccccccccc@{}}
\toprule
Ensemble & $m_{D}$ (GeV) & $m_{D^\ast}$ (GeV) & $m_{D_s}$ (GeV) & $m_{D_s^\ast}$ (GeV) & $c^2_D$ & $c^2_{D^\ast}$ & $c^2_{D_s}$ & $c^2_{D_s^\ast}$\\
\midrule
F32P21 & 1.9037(59) & 2.0288(51) & 1.9838(40) & 2.0962(38) & 1.028(27) & 1.040(32)& 1.019(19) & 1.031(22) \\
C32P23 & 1.9215(42) & 2.0139(49) & 1.9962(32) & 2.0866(41) &1.053 38)  & 1.066(48)&  1.051(28) & 1.059(34) \\
H48P32 & 1.9066(47) & 2.0354(81) & 1.9951(48) & 2.1306(70)   & 0.993(25)  & 1.028(35) & 0.994(16) & 1.030 (22)  \\
F32P30 & 1.9166(40) & 2.0362(51) & 1.9878(30) & 2.1022(38) & 0.996(28) & 0.955(32) & 1.017(19) & 0.981(22)\\
C24P29 & 1.9326(31) & 2.0299(35) & 1.9997(24) & 2.0904(44)   & 1.041(31) & 1.035(58) & 0.989(23)  & 1.019(30) \\
C48P14 & 1.8995(81) & 1.979(17) & 2.0037(36) & 2.0860(63)   & 0.90(12) & 0.92(18)& 0.961(55) & 0.966(86) \\
Extrapolation & 1.8691(86) & 2.0235(91) & 1.9794(61) & 2.1124(78) &   \\
PDG & 1.86484(5) & 2.00685(5) & 1.96835(7) & 2.1066(34)   \\
\bottomrule
\end{tabular}
\caption{This table summarizes the meson masses and the fitted $c^2$ values for each ensemble. After combined chiral and continuum extrapolations, the final spectrum results only differ from the PDG values by less than 2 sigmas.}
\label{tab:pole_mass_params}
\end{table*}

Since the charm quark is involved, we employ 
the lattice version of the dispersion relation~\cite{Bhattacharya:1995fz}
\begin{equation}
4 \sinh^2\left(\frac{aE_n}{2}\right) = 4 \sinh^2\left(\frac{am}{2}\right) + 4c^2 \sin^2\left(\frac{a\vec{p}_n}{2}\right)
\label{eq11}
\end{equation}
to fit our data.
The dispersion relation results for the $D$ meson on the F32P21 ensemble are shown in Fig.~\ref{fig2} as an example. The data points are obtained from the fits detailed in Fig.~\ref{fig1}. The slope, which corresponds to $c^2$, is consistent with one within 2 sigmas. 
Based on the discussion in previous lattice studies, e.g., Ref.~\cite{Becirevic:2014rda}, it is also beneficial to use this lattice form of momenta when determining the $Q^2$ for effective form factors:
\begin{align}
Q^2&=\left(2\sinh\left(\frac{aE_f}{2}\right)-2\sinh\left(\frac{aE_i}{2}\right)\right)^2\\ \nonumber
&-\left(2\sin\left(\frac{a\vec{p}_f}{2}\right)-2\sin\left(\frac{a\vec{p}_i}{2}\right)\right)^2.
\label{eqQ2}
\end{align}

The values of the fitted meson masses and the $c^2$ determined using Eq.~(\ref{eq11}) for each ensemble are listed in Table~\ref{tab:pole_mass_params} for the reader's reference. After performing joint continuum and chiral extrapolations using
\begin{equation}
m\left(a^2, m_\pi^2\right) = m\left(0, m_{\pi,\text{phy}}^2\right) + A a^2 + B \left(m_\pi^2 - m_{\pi,\text{phy}}^2\right),
\label{eq12}
\end{equation}
the final results of the meson masses differ from the PDG values by less than 2 sigma.

\subsection{Joint fit} \label{sec32}

According to Eqs.~(\ref{eq3-1},\ref{eq3-2},\ref{eq3}), the correlation functions contain a sum of matrix elements and kinematic factors of all hadron states satisfying the quantum numbers of the interpolating operators. Ideally, as shown above for the two-point function case, a clear ground-state plateau can be obtained at sufficiently large time separations $t$. However, the signal decays exponentially as the sink time $t$ increases, which thereby limits the usable time ranges. When $t$ is not large enough, e.g., in the case of three-point functions using the sink-sequential trick, contributions from excited states must be explicitly taken into account.

We select 3 different source-sink time separations in each ensemble. After excluding a few data points at both ends, the remaining segments of the three-point correlation functions are assumed to be dominated by the ground state and the first excited state. Expanding Eqs.~(\ref{eq3-1},\ref{eq3-2},\ref{eq3}) up to the first excited state yields
\begin{equation}
   C_{2i}(t) = \left(1 + \dfrac{p_i^2}{m^2_V}\right) \frac{A_i^VA_i^{V\ast}}{2E_V}e^{-E_{V}t}
+ \left(1 + \dfrac{p_i^2}{m^2_{V'}}\right) \frac{A_i^{V'} A_i^{V'\ast}}{2E_{V'}}e^{-E_{V^\prime}t},
\end{equation}
\begin{equation}
C_{2P}(t) = \frac{A^PA^{P\ast}}{2E_P}e^{-E_{P}t} + \frac{A^{P'} A^{P'\ast}}{2E_{P'}}e^{-E_{P^\prime}t},
\end{equation}
and
\begin{align}
C_{3\mu i}(t,\tau) &=\frac{A^P A_i^{V\ast} \left\langle P \left| J_\mu \right| V_{i} \right\rangle e^{-E_{P} t} e^{-\left(E_{V}-E_{P}\right) \tau}}{2E_V 2E_P} \nonumber \\
&+\frac{A^{P'} A_i^{V\ast} \left\langle P'\left| J_\mu \right| V_{i} \right\rangle e^{-E^\prime_P t} e^{-\left(E_{V}-E^\prime_P\right) \tau}}{2E_V 2E^\prime_P} \nonumber \\
&+\frac{A^P A_i^{V'\ast} \left\langle P \left| J_\mu \right| V'_{i} \right\rangle e^{-E_{P} t} e^{-\left(E^\prime_V -E_P\right) \tau}}{2E^\prime_V 2E_P} \nonumber \\
&+\frac{A^{P'} A_i^{V'\ast} \left\langle P' \left| J_\mu \right| V'_{i} \right\rangle e^{-E^\prime_P t} e^{-\left(E^\prime_V -E^\prime_P\right) \tau}}{2E^\prime_V 2E^\prime_P},
\label{eq14}
\end{align}
where $V$ denotes the ground vector states, $V'$ indicates the first excited states, and $P$ and $P'$ are the same for the pseudoscalar states. Since many parameters are shared in both the two-point and three-point functions, we perform a joint fit for them and account for the correlations among all the data points of the two-point and three-point correlators through a large covariance matrix.  To address potential sensitivities to small singular values arising from strong correlations, we investigated the effect of Singular Value Decomposition (SVD) truncations. As demonstrated in Table VI of the Appendix, using the fits from the F32P21 ensemble as a representative case, the results remain stable for cuts ranging from $10^{-12}$ to $10^{-6}$. This stability indicates the absence of anomalously small singular values that could distort our physical observables. Empirically, we adopt the results obtained with an SVD cut of $10^{-12}$ as our final values. 

The matrix elements $\left\langle P \left| J_\mu \right| V_{i} \right\rangle$ obtained through the joint fits can be used to calculate the effective form factor $V_{\text{eff}}$ mentioned in Eq.~(\ref{eq4}). For the cases studied in this work, the explicit forms are given by
\begin{align}
    V_{\text{eff}}^{D^{\ast+}D^+\gamma} &=-\frac{1}{3}V_l^{{D^{\ast+}D^{+}}\gamma}+\frac{2}{3}V_c^{{D^{\ast+}D^{+}}\gamma},\nonumber \\
    V_{\text{eff}}^{D^{\ast 0}D^0\gamma} &=\frac{2}{3}V_l^{D^{\ast0}D^{0}\gamma}+\frac{2}{3}V_c^{D^{\ast0}D^{0}\gamma},\nonumber \\
    V_{\text{eff}}^{D_s^{\ast +}D^+_s\gamma} &=-\frac{1}{3}V_s^{D_s^{\ast +}D^+_s\gamma}+\frac{2}{3}V_c^{D_s^{\ast +}D^+_s\gamma}, 
    \label{eq15}
\end{align}
where $V_l^{{D^{\ast+}D^{+}}\gamma}$ corresponds to the matrix element $\left\langle D^+ \left| \bar{l}\gamma_\mu l \right| D^{\ast +} _{i} \right\rangle$, and the same naming logic applies to other notations. The relative sign in Eq.~(\ref{eq15}) arises from the spin flip involved in the radiative decay from vector to pseudoscalar~\cite{Donald:2013sra}. 

\begin{figure}[htbp]  
\centering
    \includegraphics[width=0.45\textwidth]{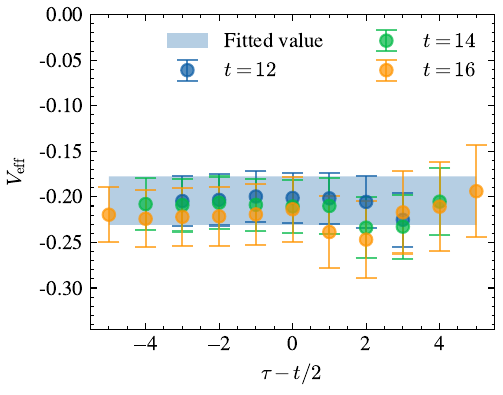}
    \hspace{0.5em}
    \includegraphics[width=0.45\textwidth]{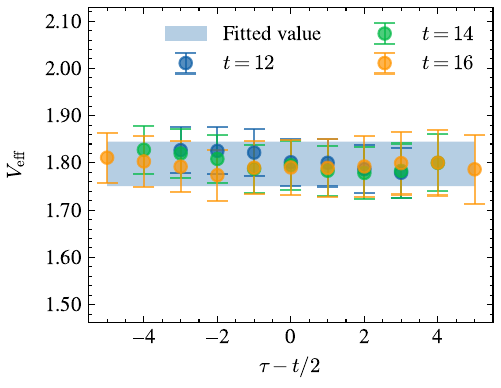}
   \includegraphics[width=0.45\textwidth]{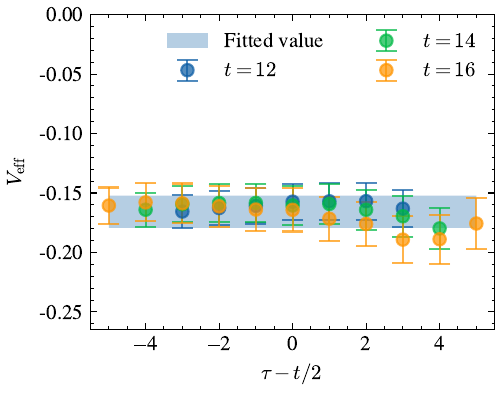}
   \caption{Results for $V_{\text{eff}}^{D^{\ast+}D^+\gamma}$, $V_{\text{eff}}^{D^{\ast0}D^0\gamma} $ at $\left(Qa\right)^2=0.0670$ and $V_{\text{eff}}^{D_s^{\ast +}D^+_s\gamma}$ at $\left(Qa\right)^2=0.0681$ on the F32P21 ensemble. The blue bands represent the matrix elements obtained from the fits, while the data points depict the values after subtracting the excited-state contributions from original three-point function data points.}\label{fig:3}
\end{figure}

In Fig.~\ref{fig:3}, we show the fitted results of the effective form factors $V_{\text{eff}}$ on the F32P21 ensemble as an example. The data points in different colors represent the values of $V_{\text{eff}}$ converted from the original data points at different source-sink time separations $t$. Specifically, they are obtained by subtracting all the excited-state contributions (the last 3 terms as formulated in Eq.~(\ref{eq14})) and removing the extra coefficients in the first term of Eq.~(\ref{eq14}), using the joint-fit results. Thus, their values correspond to the desired matrix elements.
In the plots, clear plateaus are observed, demonstrating that the contributions from the  excited states have been properly removed via our joint fits. The blue bands show the fitted results of $V_{\text{eff}}$ from the joint fits, whose central values and uncertainties agree well with the subtracted data points.

\begin{figure}[htbp]
    \centering    
    \includegraphics[width=0.45\textwidth]{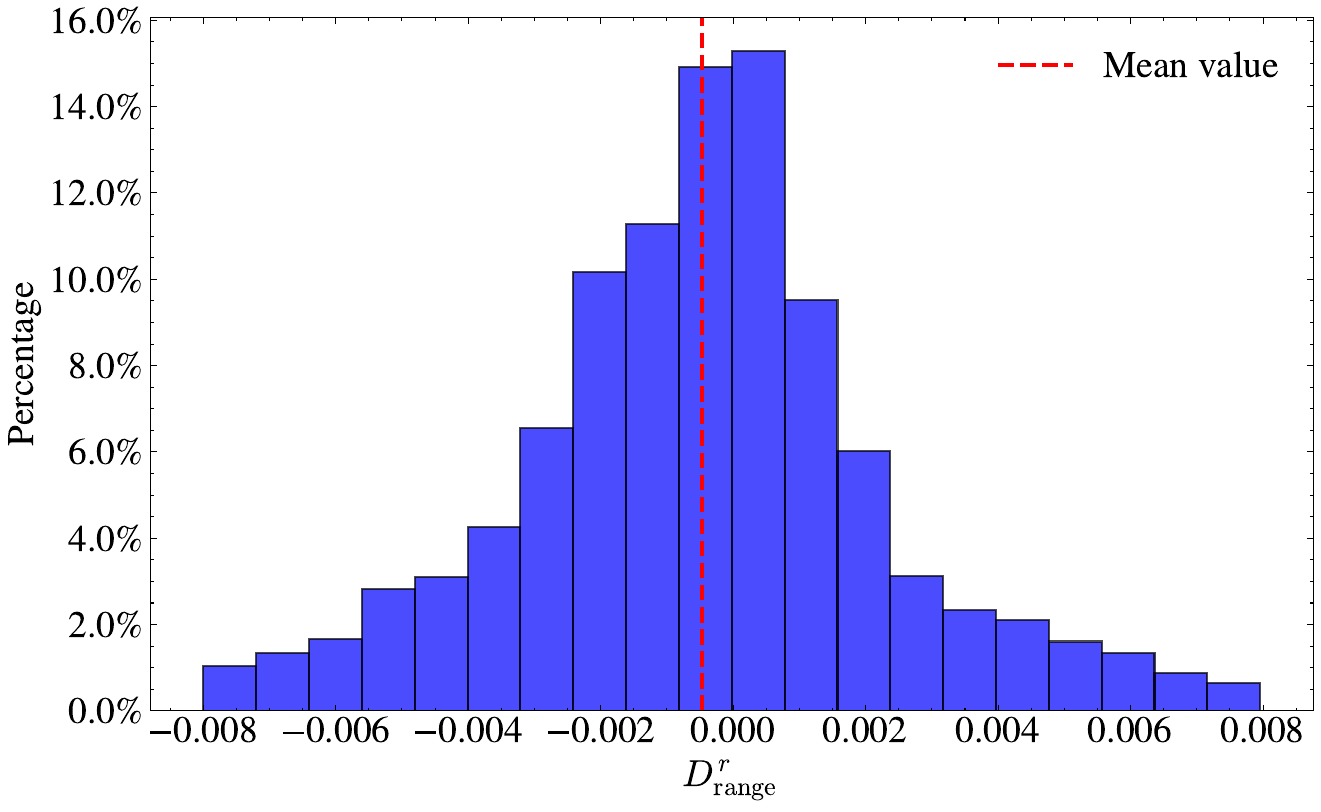}
    \includegraphics[width=0.45\textwidth]{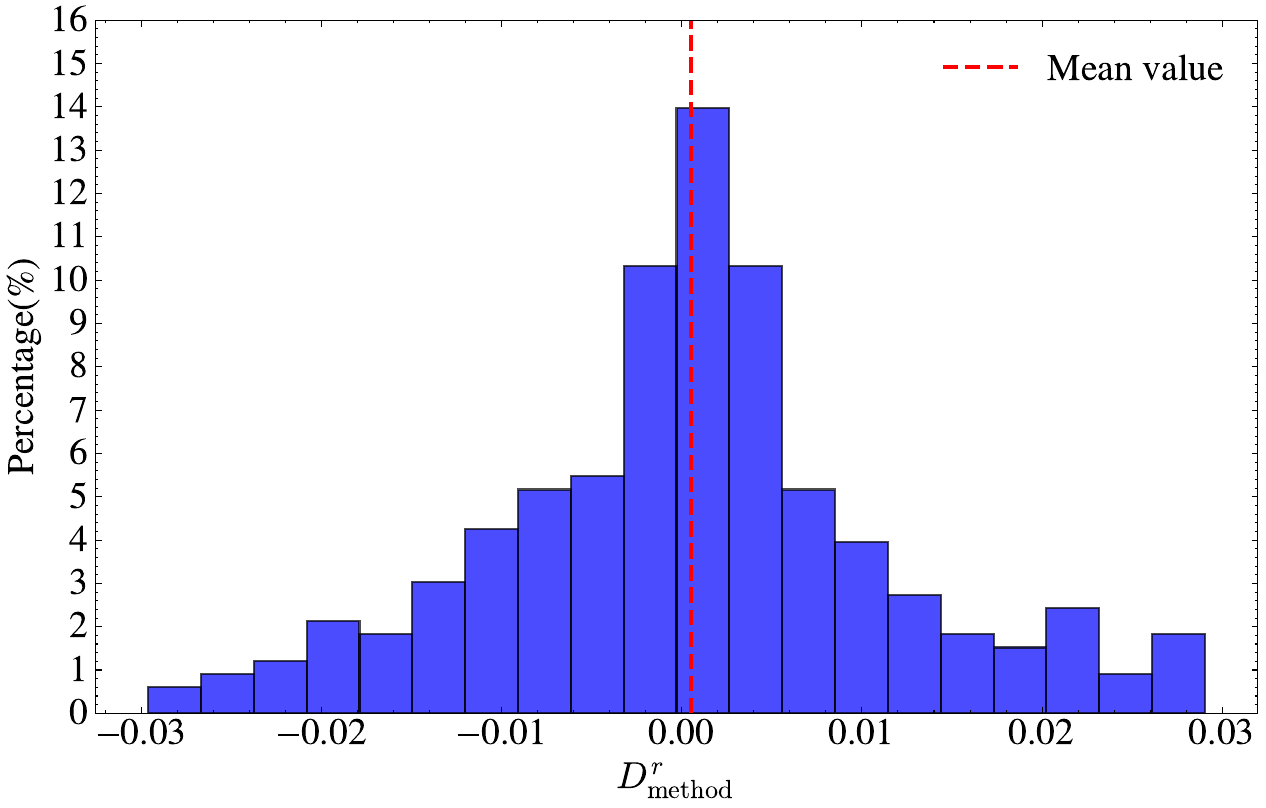} 
    \caption{The upper panel illustrates the distribution of $D^r_{\rm range}$ for the $D^{\ast +}$ meson case, which accounts for the systematic uncertainty for the choice of fit ranges, while the lower panel shows the distribution of $D^r_{\rm method}$, which accounts for the systematic uncertainty for the choice of fit methods. Both results are
    from the ensemble F32P21. See the main text for detailed discussion.
    }\label{fig:6}
\end{figure}

In the joint fits,  
we selected 48 different fit ranges for each channel, the details of which are summarized in Table~\ref{tab:fitrange}.  Most fits yield a $\chi^2/{\text{dof}} < 1.2$; they are kept as valid results. Fits with $\chi^2/{\text{dof}} > 1.2$ are discarded. Among the valid fits, the result with $\chi^2/{\text{dof}}$ closest to one is chosen as the main result, and the other valid fits are used to estimate the corresponding systematic uncertainties\cite{Liang:2023jfj}.
In explicit detail, for all the effective form factors $V_{\text{eff}}^{D^{\ast+}D^+\gamma}$, $V_{\text{eff}}^{D^{\ast0}D^0\gamma}$, and $V_{\text{eff}}^{D_s^{\ast +}D^+_s\gamma}$ with different momenta and polarizations, we calculate $D^r_{\rm range}$, which is the relative difference
between the chosen main result and the other valid fit results. The upper panel of Fig.~\ref{fig:6} shows an example of the distribution of $D^r_{\rm range}$ over the gauge configurations for $V_{\text{eff}}^{D^{\ast0}D^0\gamma}$ on the F32P21 ensemble. The distributions of $D^r_{\rm range}$ are all approximately Gaussian, with the mean value close to zero. We, therefore, take the standard deviation of these distributions as the estimation of the systematic uncertainties arising from the choice of fit ranges.   


\begin{table*}[htbp]
\centering
\setlength{\tabcolsep}{4pt}  
\begin{tabular}{@{} l *{3}{c} @{}}
\toprule
Ensemble & 2pt$_{\text{vector}}$ range & 2pt$_{\text{pseudoscalar}}$ range & 3pt function range \\
\midrule
F32P21 & $(3,19) +\text{range}(0,4)$ & $(3,19)+\text{range}(0,4)$ & $(3, t_{\text{sep}}-3)+\text{range}(-1,2)$ \\
C32P23 & $(2,18) +\text{range}(0,4)$ & $(2,18)+\text{range}(0,4)$ & $(2, t_{\text{sep}}-2)+\text{range}(-1,2)$ \\
H48P32 & $(4,20) +\text{range}(0,4)$ & $(4,20)+\text{range}(0,4)$ & $(4, t_{\text{sep}}-4)+\text{range}(-1,2)$ \\
F32P30 & $(3,19) +\text{range}(0,4)$ & $(3,19)+\text{range}(0,4)$ & $(3, t_{\text{sep}}-3)+\text{range}(-1,2)$ \\
C24P29 & $(2,18) +\text{range}(0,4)$ & $(2,18)+\text{range}(0,4)$ & $(2, t_{\text{sep}}-2)+\text{range}(-1,2)$ \\
C48P14 & $(2,18) +\text{range}(0,4)$ & $(2,18)+\text{range}(0,4)$ & $(2, t_{\text{sep}}-2)+\text{range}(-1,2)$ \\
\bottomrule
\end{tabular}
  \caption{The columns of $2\text{pt}_{\text{vector}}$ and $2\text{pt}_{\text{pseudoscalar}}$ collect the fit ranges of two-point functions. The notation ``($t_s$,$t_e$)+range(0,4)'' indicates that 
  the initial fit range is ($t_s$,$t_e$) and we shift both the start and end points of the range forward by 0, 1, 2, and 3 lattice spacings while keeping the total fit length constant. For two-point functions with non-zero momenta, we empirically shift the fit ranges backward by approximately $p^2/2$ lattice spacings. This adjustment ensures that the majority of the fits yield $\chi^2/{\rm dof}<1.2$. For the three-point functions, we have $t_{\rm sep} = 10, 12, 14$ for the C ensembles, $t_{\rm sep} = 12, 14, 16$ for the F ensembles and $t_{\rm sep} = 16, 18, 20$ for the H ensemble. The notation ``($t_c$,$t_{\rm sep}-t_c$)+range($-1$,$2$)'' means for each $t_{\rm sep}$ the initial fit range is ($t_c$,$t_{\rm sep}-t_c$) and we  change $t_c$ forward or backward by one lattice spacing. }
\label{tab:fitrange}
\end{table*}

\begin{figure}[htbp]
    \centering
    \includegraphics[width=0.45\textwidth]{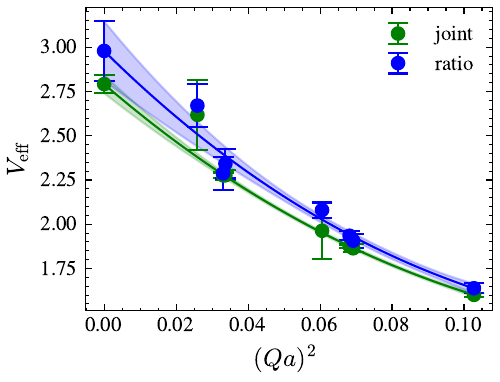}
    \caption{Comparison of the two different fit methods on the F32P21 ensemble for $V_{\text{eff}}^{D^{\ast0}D^0\gamma}$ with different momentum transfers. The blue band and data points represent the results from the two-state fit method, while the green ones are from the joint fit method.}
    \label{fig:5}
\end{figure}

The other commonly used method to extract the matrix elements is called the two-state fit, which
handles the three-point to two-point function ratios. Considering the ground state and the first excited state, the ratios can be expressed as:
\begin{align}
R_{\mu i} &= C_{3 \mu i} (t, \tau) \sqrt{\frac{2 E^{P} C_{2}^{P} (\tau)}{C_{2}^{P} (t - \tau) C_{2}^{P} (t)}}
   \sqrt{\frac{2 E^{V} C_{2}^{V} (t -\tau)}{C_{2}^{V} (\tau) C_{2}^{V} (t)}} \nonumber \\
&= \frac{\left\langle P \left| j_{\mu} \right| V_i \right\rangle}{\sqrt{1 + \frac{{p_i^V}^2}{m^2_V}}} + C_1 e^{-\delta E_1\tau} + C_2 e^{-\delta E_2(t-\tau)}+C_3 e^{-\delta E_3t}.
\label{eq16}
\end{align} 
We also perform two-state fits to our data and check the difference between the two-state fit results and those from the joint fits. 
The fit results of the two methods on the F32P21 ensemble for $V_{\text{eff}}^{D^{\ast0}D^0\gamma}$ are presented in Fig.~\ref{fig:5}, serving as an example of comparison. Our tests show that the central values are consistent within errors, while the joint fits yield higher signal-to-noise ratios. Additionally, the parameters of joint fits are more interpretable, as Eq.~(\ref{eq14}) is explicit. In contrast, Eq.~(\ref{eq16}) assumes no excited states in the two-point functions, which is less reasonable. For the reasons above, the joint fit method is chosen to obtain the main results reported in this work.

Following a similar procedure for estimating the systematic uncertainties from the choice of fit ranges, we also calculate $D^r_{\rm method}$, which is the relative difference
between the form factor results from the two fit methods. 
We gather $D^r_{\rm method}$ for all three channels $V_{\text{eff}}^{D^{\ast+}D^+\gamma}$, $V_{\text{eff}}^{D^{\ast0}D^0\gamma}$, and $V_{\text{eff}}^{D_s^{\ast +}D^+_s\gamma}$ 
with different momenta and polarizations together,
and the corresponding distribution
of $D^r_{\rm method}$ over the gauge configurations on the F32P21 ensemble is plotted in the lower panel of Fig.~\ref{fig:6} as an example. Again, the standard deviation of this distribution is taken as the systematic uncertainty arising from the choice of fit methods. Detailed fit results on each ensemble are listed in the Appendix.

\subsection{Momentum transfer extrapolation}\label{sec33}

\begin{figure}[htbp]
    \centering   
    \includegraphics[width=0.42\textwidth]{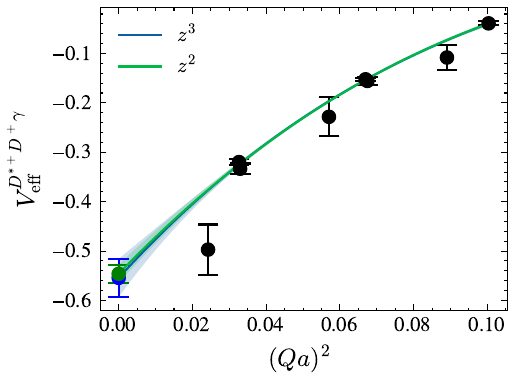}
    \includegraphics[width=0.42\textwidth]{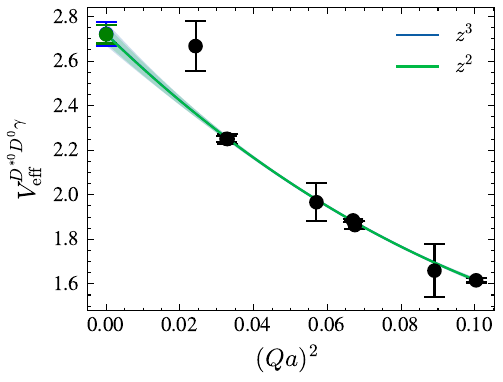}
    \includegraphics[width=0.42\textwidth]{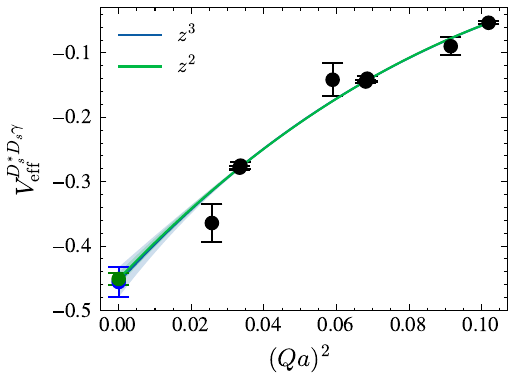} 
    \caption{The momentum transfer extrapolation of $V_{\text{eff}}^{D^{\ast+}D^+\gamma}$, $V_{\text{eff}}^{D^{\ast0}D^0\gamma}$, and $V_{\text{eff}}^{D_s^{\ast +}D^+_s\gamma}$ on the F32P21 ensemble are shown from top to bottom. The blue band and the blue point at $\left(Qa\right)^2 = 0$ correspond to the fit result with truncation at $z^3$, while the green ones are for the fit result with truncation at $z^2$. The difference between the values at $\left(Qa\right)^2 = 0$ are taken as the systematic uncertainty.}\label{fig:4}
\end{figure}

The matrix elements at the same $Q^2$ but with different momentum setups are all computed and averaged. We then employ the model-independent $z$-expansion to fit the data and extract the $V_{\text{eff}}(0)$. The explicit form is as follows~\cite{Hill:2010yb}:
\begin{equation}
z\left(t,t_{\mathrm{cut}},t_{0}\right) = \frac{\sqrt{t_{\mathrm{cut}}-Q^2} - \sqrt{t_{\mathrm{cut}}-t_{0}}}{\sqrt{t_{\mathrm{cut}}-Q^2} + \sqrt{t_{\mathrm{cut}}-t_{0}}},
\end{equation}
\begin{equation}
V_{\mathrm{eff}}\left(Q^2\right) = \sum_{i=0}^n a_i z^i,
\end{equation}
where $t_{\mathrm{cut}}=\left(m_{\pi}+m_P\right)^2$ and $t_{0}$ are free parameters. A commonly used choice is $t_{0}^{\mathrm{opt}}=t_{\mathrm{cut}}\left(1-\sqrt{1+Q^{2}_{\max}/t_{\mathrm{cut}}}\right)$~\cite{Hill:2010yb}. 
In the $z$-expansion fits, truncating the series at either the second or third order leads to differences in the fitted central values.
An example of the F32P21 ensemble is presented in Fig.~\ref{fig:4}, where the green point at $\left(Qa\right)^2=0$ corresponds to the extrapolated result with truncation up to $z^2$ terms, while the blue point is for the result with truncation up to $z^3$ terms. The results for the other ensembles are listed in the Appendix.
We find that the $z^3$ terms are not statistically significant. Thus, we take the extrapolated values of the $z^2$ expansions to be the main results and
consider the difference between the $z^2$ and $z^3$
truncated results as the systematic uncertainties arising from the momentum-transfer extrapolations.

\subsection{Continuum and chiral extrapolation}\label{sec34}

After obtaining the effective form factors $V_{\mathrm{eff}}$ at $Q^2 = 0$ for each ensemble and incorporating the systematic uncertainties from the choice of fit ranges, fit methods, and truncations in the momentum transform extrapolations, we convert the results of $V_{\mathrm{eff}}(0)$ to the coupling constant via
\begin{equation}
    g_{VP}=\frac{2V_{\text{eff}}(0)}{m_V+m_P}.
\end{equation}
We then perform a combined continuum and chiral extrapolation with respect to $a^2$ and $m_\pi^2$ using the following empirical form:
\begin{equation}
g_{VP}\left(a^2, m_\pi^2\right) = g_{VP}\left(0, m_{\pi,\text{phy}}^2\right) + A a^2 + B\left(m_\pi^2 - m_{\pi,\text{phy}}^2\right),
\label{eq19}
\end{equation}
where $A$ and $B$ are free parameters. When performing the extrapolation using Eq.~(\ref{eq19}), for $g_{D^{\ast+} D^+ \gamma}$ and $g_{D_s^{\ast +} D^+_s \gamma}$, we find $\chi^2/\text{dof} < 0.7$, confirming that the extrapolation form describes our data well. In contrast, the fit for $g_{D^{\ast0} D^0 \gamma}$ gives $\chi^2/\text{dof} = 1.7$. 
We also use a modified form with a logarithmic term inspired by chiral perturbation theory
\begin{equation}
\begin{aligned}
g_{VP}\left(a^2, m_\pi^2\right) &= g_{VP}\left(0, m_{\pi,\text{phy}}^2\right) + A a^2 + B\left(m_\pi^2 - m_{\pi,\text{phy}}^2\right) \\
&+ C \left( m_\pi^2\log\left(m_\pi^2\right) - m_{\pi,\text{phy}}^2\log\left(m_{\pi,\text{phy}}^2\right) \right).
\label{eq20}
\end{aligned}
\end{equation}

  The feasibility of performing such an extrapolation for $g_{D^{\ast+} D^+ \gamma}$ and $g_{D_s^{\ast +} D^+_s \gamma}$ stems from the inclusion of the C48P14 ensemble, which imposes a stringent constraint on the chiral logarithm term. As illustrated in Fig.~\ref{fig:coupling_compare} of the Appendix, the addition of this physical-mass ensemble leads to a substantial reduction in the  uncertainties when the chiral logarithm term is considered. Furthermore, by comparing the results obtained with and without these terms, we are able to perform a rigorous evaluation of the systematic uncertainty associated with the chiral extrapolation. However, for $g_{D^{\ast0} D^0 \gamma}$ , the $\chi^2/{\text{dof}}$ of this form is still around 1.4, which remains relatively high for a fit with only a few degrees of freedom.

\begin{figure*}[htbp]
    \centering
    \includegraphics[width=0.24\textwidth]{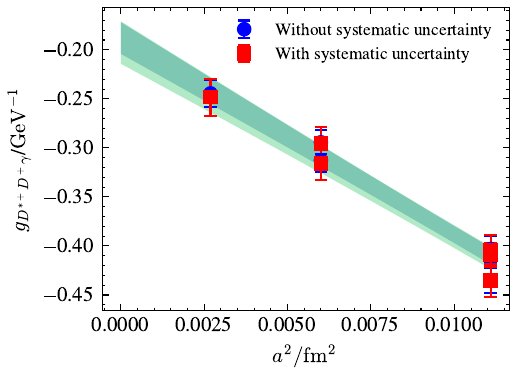}
    \includegraphics[width=0.24\textwidth]{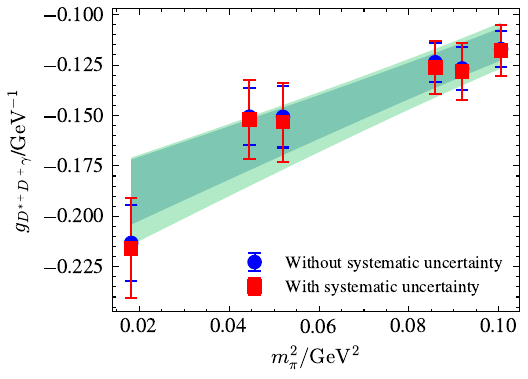}
    \includegraphics[width=0.24\textwidth]{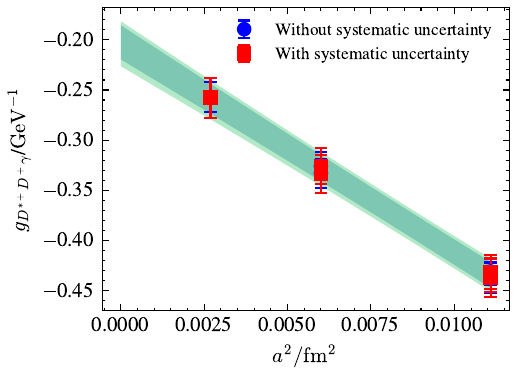}
    \includegraphics[width=0.24\textwidth]{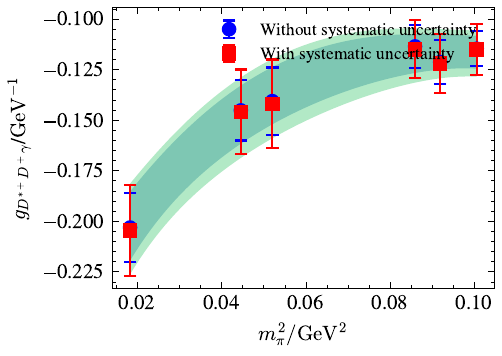}
    
    \includegraphics[width=0.24\textwidth]{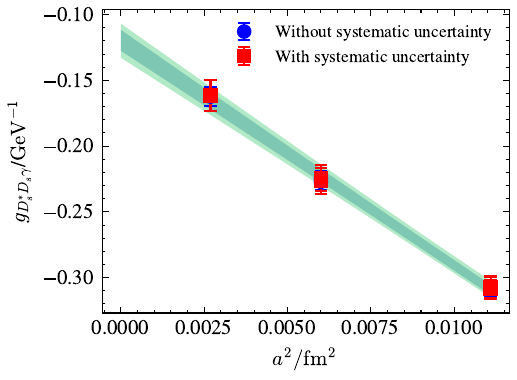}
    \includegraphics[width=0.24\textwidth]{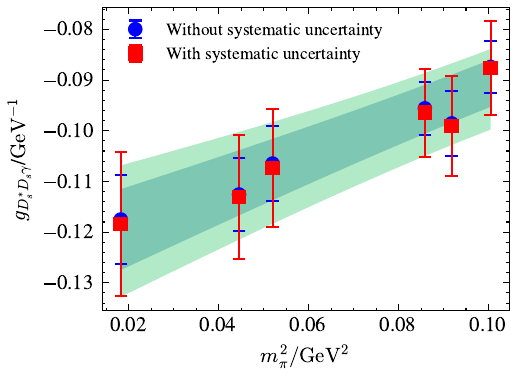}
    \includegraphics[width=0.24\textwidth]{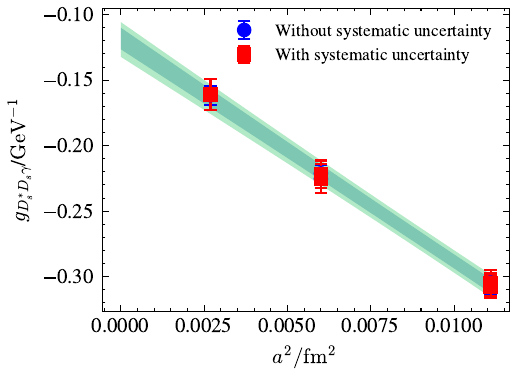}
    \includegraphics[width=0.24\textwidth]{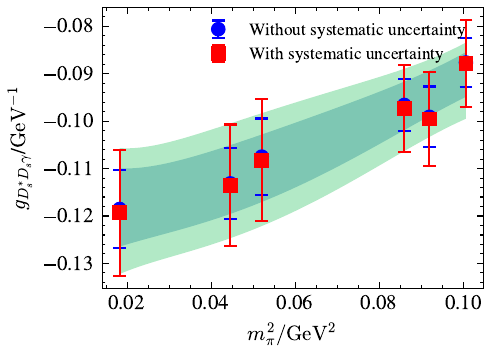}
    
    \includegraphics[width=0.24\textwidth]{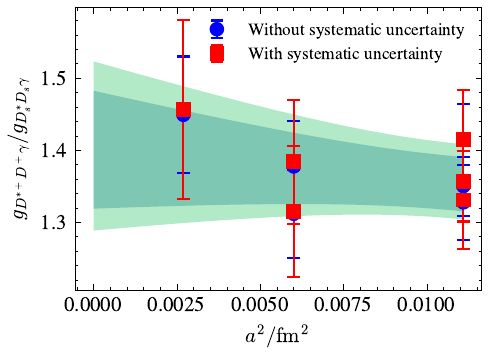}
    \includegraphics[width=0.24\textwidth]{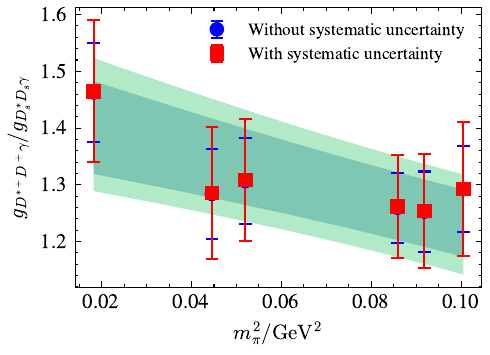}
    \includegraphics[width=0.24\textwidth]{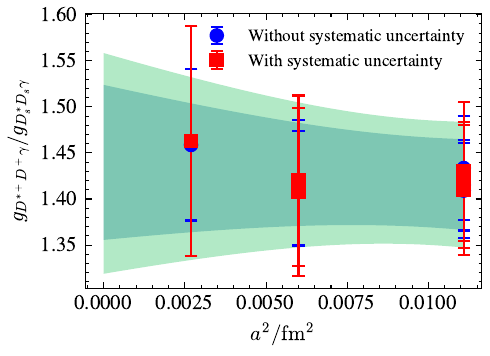}
    \includegraphics[width=0.24\textwidth]{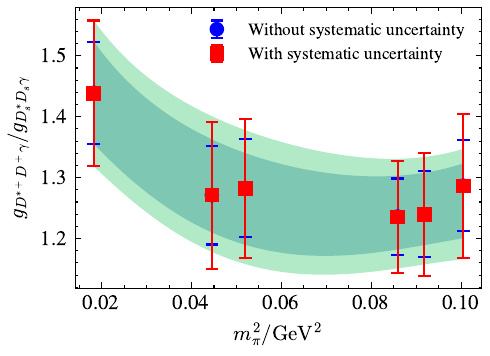} 

    \includegraphics[width=0.24\textwidth]{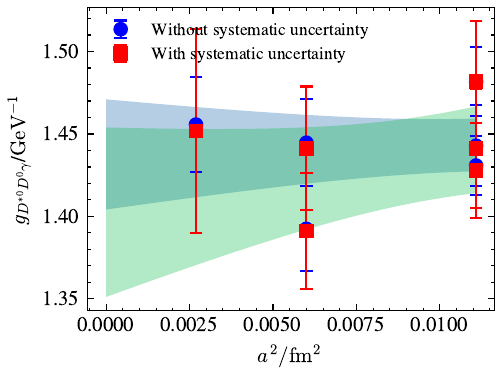}
    \includegraphics[width=0.24\textwidth]{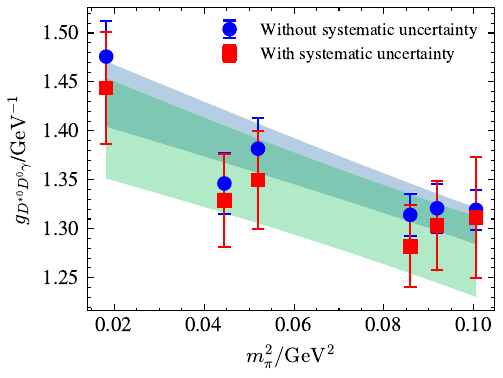}
    \includegraphics[width=0.24\textwidth]{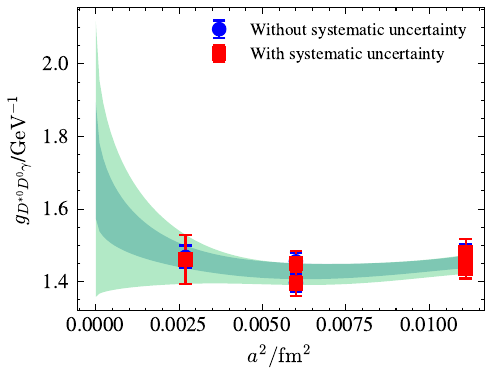}
    \includegraphics[width=0.24\textwidth]{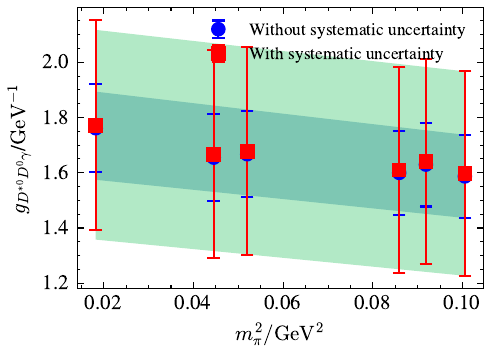}
    \caption{The first three rows display the combined continuum and chiral extrapolation for \( g_{D^{\ast+} D^+ \gamma} \), $g_{D_s^{\ast +} D^+_s \gamma}$, and $g_{D^{\ast+} D^+ \gamma}/g_{D_s^{\ast} D_s \gamma}$, respectively. The two left columns show the dependence on lattice spacing and pion mass based on fit with Eq.~(\ref{eq19}), while the two right columns present the results obtained from Eq.~(\ref{eq20}). Both formulas yield $\chi^2/\text{dof}$ below 0.3. The differences between the two formulas are taken as the corresponding systematic uncertainties. The fourth row shows the results for \( g_{D^{\ast0} D^0 \gamma} \). Again, the left two plots correspond to Eq.~(\ref{eq19}), which fails to produce an acceptable $\chi^2/\text{dof}$, whereas the right two plots show the results using Eq.~(\ref{eq21}), which describes well the lattice data.}
    \label{result_a}
\end{figure*}

We attribute this observation to discretization effects. In fact, when calculating vector current matrix elements using clover fermions, improvements to the current operator should be considered; otherwise, $O(a)$ discretization errors may be significant~\cite{Becirevic:2009xp,Sint:1997jx,Gerardin:2018kpy,Heitger:2020zaq}.
An explicit form is as follows:
\begin{equation}
    V_{\mu}^{\text{im}}(x) = Z_V\left(g_0^2\right) \left(1 + b_V\left(g_0^2\right)\left(am_q\right)\right) \left[V_{\mu}(x) + c_V \partial_{\nu} T_{\mu\nu}(x)\right],
\end{equation}
where $b_V$ and $c_V$ are the improvement coefficients. Therefore, we try to use a form with additional $a$ dependence to fit our data
\begin{equation}
g_{VP}\left(a^2, m_\pi^2\right) = g_{VP}\left(0, m_{\pi,\text{phy}}^2\right) + A a^2 + B\left(m_\pi^2 - m_{\pi,\text{phy}}^2\right)+ Ca.
\label{eq21}
\end{equation}
This time, for $g_{D^{\ast0} D^0 \gamma}$, the $\chi^2/{\text{dof}} = 0.2$, and the coefficient $C$ has a signal-to-noise ratio larger than 3, showing clear $O(a)$ effects.
Therefore, the results using Eq.~(\ref{eq21}) are adopted for the $g_{D^{\ast0} D^0 \gamma}$ case.
At the same time,
for $g_{D^{\ast+} D^+ \gamma}$ and $g_{D_s^{\ast +} D^+_s \gamma}$, the fits yield statistically zero results for the $C$ coefficients, and the central values are consistent with the previous results within errors, which indicates that there are no significant $O(a)$ effects for the positively charged cases at the current statistical precision. All these extrapolations are shown in Fig.~\ref{result_a}.

In addition, we carry out a check on different methods to incorporate 
the vector renormalization constants. The constants listed in Table.~\ref{tab:lattice_params} are the ones at the chiral limit. We also use those directly obtained from each gauge ensemble. After the combined chiral and continuum extrapolation described above, especially with the $O(a)$ term included for the neutral case, both methods yield consistent results, which are formally stated in Eq.~(\ref{eq21}).

One could also compute the ratio \( g_{D^{\ast +} D^+ \gamma} / g_{D_s^{\ast} D_s \gamma} \) on each ensemble and then perform the extrapolation using Eq.~(\ref{eq19}) and Eq.~(\ref{eq20}). In this way, we obtain
\[
\frac{g_{D^{\ast +} D^+ \gamma}}{g_{D_s^{\ast +} D^+_s \gamma}} = 1.43(12).
\]
This value is important for understanding the $\rm SU(3)$ flavor symmetry and its breaking.

\section{Conclusion}\label{sec4}

This work presents the first systematic calculation of the radiative decay coupling constants of charmed mesons.
We have one ensemble at the physical pion mass \( m_{\pi} = 135\ \text{MeV} \) and one ensemble with lattice spacing \( a \sim 0.05\ \text{fm} \). Continuum and chiral extrapolations are carefully performed, and systematic uncertainties 
are meticulously estimated.

For the process \( D^{\ast 0} \to D^0 \gamma \), our result for \( g_{D^{\ast 0} D^0 \gamma} \) is \( 1.73(16)(34)\ \text{GeV}^{-1} \), yielding a partial width of
\[
\Gamma(D^{\ast 0} \to D^0 \gamma) = 18.2(3.4)(7.1)\ \text{keV},
\]
where the errors are the total uncertainties.
Combining this result with the experimental branching fraction \( \mathcal{B}(D^{\ast 0} \to D^0 \gamma) = (34.5 \pm 0.8 \pm 0.5)\% \), we predict the total width of the $D^{\ast 0}$ meson to be 
\[
\Gamma(D^{\ast 0}) = 52(10)(20)\ \text{keV},
\]
which is well below the experimental upper limit of \( 2.1\ \text{MeV} \).
For the process $ D^{\ast +} \to D^+ \gamma$, we find \( g_{D^{\ast +} D^+ \gamma} = 0.204(17)(14)(11)\ \text{GeV}^{-1} \), which corresponds to a radiative decay width of
\[
\Gamma(D^{\ast +} \to D^+ \gamma) = 0.253(42)(35)(27)\ \text{keV}.
\]
Similarly, using the total decay width \( \Gamma(D^{\ast +}) = 83.3 \pm 1.2 \pm 1.4\ \text{keV} \) from the BaBar collaboration in 2013 \cite{BaBar:2013thi}, we obtain the branching ratio
\[
\mathcal{B}(D^{\ast +} \to D^+ \gamma) = 0.304(51)(42)(33)\%,
\]
which is significantly lower than the experimental value $(1.68 \pm 0.42 \pm 0.29)\%$ reported by CLEO in 1998. Future experiments with improved precision should better elucidate this dependence.

For the process $D_s^{*+} \to D_s^+ \gamma$, we determine $g_{D_s^{*+} D^+_s \gamma} =-0.120(9)(10)(1) \text{ GeV}^{-1}$ and predict the radiative decay width
\begin{equation}
\Gamma(D_s^{*+} \to D_s^+ \gamma) = 0.094(14)(16)(2) \text{ keV}.\nonumber
\end{equation}
Assuming that the decays of $D^{\ast +} _s$ are entirely accounted for by strong and radiative decays, the total width of $D^{\ast +} _s$ can be estimated as
\[
\Gamma(D_s^{\ast +} ) = 0.100(16)(17)(5)\ \text{keV}.
\]
Besides, we calculate the ratio \( g_{D^{\ast +} D^+ \gamma} / g_{D_s^{\ast} D_s \gamma} = 1.43(8)(9)(3) \). This result serves as an input for related phenomenological models. We list our results on each ensemble, the error budget, and the final extrapolated results with detailed systematic uncertainties in Table~\ref{tab:budget} for the reader's reference. A comparison with other works is summarized in Table~\ref{tab_result}.

Studies that include disconnected contributions and carefully treat finite-volume effects are planned for the near future.  The significant discrepancy between our lattice prediction for $\mathcal{B}(D^{\ast+}\to D^{+}\gamma)$ and the 1998 CLEO result calls for a renewed experimental investigation. In the near future, we expect this situation to be clarified by several high-intensity experiments. The BESIII collaboration, having recently accumulated a massive 20 fb$^{-1}$ data sample at the $D\bar{D}$ threshold, is currently analyzing charmed meson decays with unprecedented precision. Simultaneously, Belle II is ramping up its luminosity, and its recent discovery of the $D_{s0}^\ast(2317)^+ \to D_s^{\ast+} \gamma$ transition underscores its potential to measure rare radiative mode~\cite{Belle-II:2025dzk}. These upcoming results will provide a critical test of our first-principles calculations and may resolve the long-standing tension in the $D^\ast$ radiative sector. 

\begin{table*}[htbp]
    \centering
    \begin{tabular}{ccccccc}
        Ensemble & F32P21 & C32P23 & H48P32 & F32P30 & C24P29 & C48P14 \\
        \midrule
        \multicolumn{7}{c}{$g_{D^{\ast+}D^+\gamma}$} \\
        \midrule
        Value without systematic (GeV$^{-1}$) & -0.2709(91) & -0.373(11) & -0.1710(68) & -0.2470(75) & -0.3459(84) & -0.435(13) \\
        Momentum transfer extrapolation (GeV$^{-1}$) & 0.008 & 0.0001 & 0.0003 & 0.002 & 0.002 & 0.000005 \\
        Fit range (GeV$^{-1}$) & 0.0046 & 0.0053 & 0.0025 & 0.0047 & 0.0039 & 0.0064 \\
        Fit method (\%) & 1.11 & 1.38 & 4.00 & 1.50 & 1.01 & 1.84 \\
        Value with systematic (GeV$^{-1}$) & -0.271(13) & -0.373(13) & -0.1710(99) & -0.2470(98) & -0.346(10) & -0.435(17) \\
        \multicolumn{1}{c}{Extrapolated value with Eq.~(\ref{eq19})} & \multicolumn{6}{c}{$-0.193(16)(15)$} \\
        \multicolumn{1}{c}{Extrapolated value with Eq.~(\ref{eq20})} & \multicolumn{6}{c}{$-0.204(17)(14)$} \\
        \multicolumn{1}{c}{Final result} & \multicolumn{6}{c}{$-0.204(17)(14)(11)$} \\
        \midrule
        \multicolumn{7}{c}{$g_{D_s^{\ast +} D^+_s \gamma}$} \\
        \midrule
        Value without systematic (GeV$^{-1}$) & -0.2153(46) & -0.2963(50) & -0.1335(41) & -0.2013(49) & -0.2854(40) & -0.3074(55) \\
        Momentum transfer extrapolation (GeV$^{-1}$) & 0.005 & 0.00007 & 0.002 & 0.0009 & 0.0004 & 0.000004 \\
        Fit range (GeV$^{-1}$) & 0.0037 & 0.0028 & 0.0026 & 0.0037 & 0.0030 & 0.0037 \\
        Fit method (\%) & 1.11 & 1.38 & 4.00 & 1.50 & 1.01 & 1.84 \\
        Value with systematic (GeV$^{-1}$) & -0.2153(81) & -0.2963(71) & -0.1335(75) & -0.2013(69) & -0.2854(58) & -0.3074(87) \\
        \multicolumn{1}{c}{Extrapolated value with Eq.~(\ref{eq19})} & \multicolumn{6}{c}{$-0.120(8)(10)$} \\
        \multicolumn{1}{c}{Extrapolated value with Eq.~(\ref{eq20})} & \multicolumn{6}{c}{$-0.119(8)(10)$} \\
        \multicolumn{1}{c}{Final result} & \multicolumn{6}{c}{$-0.120(9)(10)(1)$} \\
        \midrule
        \multicolumn{7}{c}{$g_{D^{\ast0}D^0\gamma}$} \\
        \midrule
        Value without systematic (GeV$^{-1}$) & 1.349(20) & 1.388(22) & 1.321(16) & 1.324(19) & 1.320(17) & 1.482(21) \\
        Momentum transfer extrapolation (GeV$^{-1}$) & 0.002 & 0.00009 & 0.002 & 0.0005 & 0.0006 & 0.000003 \\
        Fit range (GeV$^{-1}$) & 0.0097 & 0.0083 & 0.0050 & 0.0097 & 0.0093 & 0.0126 \\
        Fit method (\%) & 1.11 & 1.38 & 4.00 & 1.50 & 1.01 & 1.84 \\
        Value with systematic (GeV$^{-1}$) & 1.349(26) & 1.388(29) & 1.321(55) & 1.324(28) & 1.320(22) & 1.482(35) \\
        \multicolumn{1}{c}{Extrapolated value with Eq.~(\ref{eq21})} & \multicolumn{6}{c}{$1.73(16)(34)$} \\
        \multicolumn{1}{c}{Final result} & \multicolumn{6}{c}{$1.73(16)(34)$} \\
        \bottomrule
    \end{tabular}

    \caption{Results and systematic uncertainties on each ensemble. ``Value without systematic'' refers to results containing statistical uncertainties only. ``Momentum transfer extrapolation'', ``Fit range'' and ``Fit method'' refer to the corresponding systematic uncertainties. ``value with systematic'' indicates results that incorporate all systematic uncertainties. The difference between "Extrapolated value with Eq.~(\ref{eq19})'' and ``Extrapolated value with Eq.~(\ref{eq20})" lies in whether the logarithmic term is included in the extrapolation. The ``Final result'' is obtained by accounting for the discrepancy between these two approaches: the central values are the average of the two, and an additional systematic uncertainty is take as half of their difference. For $g_{D^{\ast 0}D^0\gamma}$, only the extrapolation using Eq.~(\ref{eq21}) yields a reasonable $\chi^2/{\text{dof}}$. Therefore, the final result is taken directly from this fit.}
    \label{tab:budget}
\end{table*}

\begin{table*}[htbp]
\centering
\renewcommand{\arraystretch}{1.5}
\begin{tabular}{lcccc}
\toprule
Ref. & \( g_{D^{\ast+}D^{+}\gamma} \) & \( g_{D^{\ast 0}D^{0}\gamma} \) & \( g_{D^{\ast}_sD^{+}_{s}\gamma} \)& \( g_{D^{\ast +}D^{+}\gamma}/g_{D^{\ast +}_sD^{+}_{s}\gamma} \) \\
\midrule
This work & \(-0.204(17)(14)(11)\) & \( 1.73(16)(34) \) & \( -0.120(9)(10)(1)\)  & \( 1.43(8)(9)(3)\) \\
\multirow{3}{*}{LQCD\cite{Becirevic:2009xp},\cite{Donald:2013sra},\cite{Meng:2024gpd},\cite{Frezzotti:2023ygt}} & \multirow{3}{*}{\(-0.2(3)\)} & \multirow{3}{*}{\(2.0(6)\)} & \(-0.087(4)\) \\
& & & \(-0.118(13)\) \\
& & & \(-0.098(19)\)   \\
\multirow{2}{*}{LCSR\cite{Li:2020rcg},\cite{Pullin:2021ebn}} & \(-0.15(11)\) & \(1.48(29)\) & \(-0.079(86)\)  \\
& \(0.40(13)\) & \(-2.11(35)\) & \(0.60(19)\)\\
HH$\chi$PT\cite{Amundson:1992yp} & \(-0.27(5)\) & \( 2.19(11) \) & \( 0.041(56) \)  \\
CCQM\cite{Tran:2023hrn}  & \(-0.45(9) \) & \(1.72(34) \) & \(-0.29(6) \)  \\
HQET+COM\cite{Cheung:2014cka} & \(-0.38(5) \) & \(1.91(9) \)   \\
Bag Model\cite{Orsland:1998de} & 0.5 & 1.1 \\
\multirow{2}{*}{RQM\cite{Jaus:1996np},\cite{Goity:2000dk}} & \(-0.30\) & \(1.85\)  \\
& \(-0.44(6)\) & \(2.15(11)\) & \(-0.19(3)\)\\

 \multirow{2}{*}{QCDSR\cite{Aliev:1994nq},\cite{Lu:2024tgy} } & \(-0.19(3) \) & \( 0.62(3) \) & \(-0.20(3) \) \\
& \(0.167(40)\) & \(0.5275(167)\) & \(0.0667(103)\)\\
\bottomrule
\end{tabular}
\caption{For the results in this work, the first uncertainty is statistical, and the second one is the systematic uncertainty from the matrix element extraction and momentum transfer extrapolation, and the last one is the systematic uncertainty from chiral extrapolation. All the couplings are in unit GeV$^{-1}$.}
\label{tab_result}
\end{table*}

\section*{Acknowledgments}
We thank Prof.\ Ying Chen for helpful discussions.
The numerical calculations are performed on the Southern Nuclear Science Computing Center (SNSC) of South China Normal University and the Xiangjiang-1 cluster at Hunan Normal University (Changsha).
This work is supported by the National Natural Science Foundation of China (NSFC) under Grants No.\ 12575085, No.\ 12175073, No.\ 12222503, No.\ 12175063, No.\ 12105108, No.\ 12205106, and No.\ 12305094.
J. L. also acknowledges the support of the Natural Science Foundation of Basic and Applied Basic Research of Guangdong Province under Grant No.\ 2023A1515012712. 
Y. M. also thanks the support from the Young Elite Scientists Sponsorship Program by Henan Association for Science and Technology with Grant No. 2025HYTP003.
\appendix
\bibliographystyle{apsrev4-2}
\bibliography{bib}

\begin{thebibliography}{27}%
\makeatletter
\providecommand \@ifxundefined [1]{%
 \@ifx{#1\undefined}
}%
\providecommand \@ifnum [1]{%
 \ifnum #1\expandafter \@firstoftwo
 \else \expandafter \@secondoftwo
 \fi
}%
\providecommand \@ifx [1]{%
 \ifx #1\expandafter \@firstoftwo
 \else \expandafter \@secondoftwo
 \fi
}%
\providecommand \natexlab [1]{#1}%
\providecommand \enquote  [1]{``#1''}%
\providecommand \bibnamefont  [1]{#1}%
\providecommand \bibfnamefont [1]{#1}%
\providecommand \citenamefont [1]{#1}%
\providecommand \href@noop [0]{\@secondoftwo}%
\providecommand \href [0]{\begingroup \@sanitize@url \@href}%
\providecommand \@href[1]{\@@startlink{#1}\@@href}%
\providecommand \@@href[1]{\endgroup#1\@@endlink}%
\providecommand \@sanitize@url [0]{\catcode `\\12\catcode `\$12\catcode `\&12\catcode `\#12\catcode `\^12\catcode `\_12\catcode `\%12\relax}%
\providecommand \@@startlink[1]{}%
\providecommand \@@endlink[0]{}%
\providecommand \url  [0]{\begingroup\@sanitize@url \@url }%
\providecommand \@url [1]{\endgroup\@href {#1}{\urlprefix }}%
\providecommand \urlprefix  [0]{URL }%
\providecommand \Eprint [0]{\href }%
\providecommand \doibase [0]{https://doi.org/}%
\providecommand \selectlanguage [0]{\@gobble}%
\providecommand \bibinfo  [0]{\@secondoftwo}%
\providecommand \bibfield  [0]{\@secondoftwo}%
\providecommand \translation [1]{[#1]}%
\providecommand \BibitemOpen [0]{}%
\providecommand \bibitemStop [0]{}%
\providecommand \bibitemNoStop [0]{.\EOS\space}%
\providecommand \EOS [0]{\spacefactor3000\relax}%
\providecommand \BibitemShut  [1]{\csname bibitem#1\endcsname}%
\let\auto@bib@innerbib\@empty
\bibitem [{\citenamefont {Bartelt}\ \emph {et~al.}(1998)\citenamefont {Bartelt} \emph {et~al.}}]{CLEO:1997rew}%
  \BibitemOpen
  \bibfield  {author} {\bibinfo {author} {\bibfnamefont {J.~E.}\ \bibnamefont {Bartelt}} \emph {et~al.} (\bibinfo {collaboration} {CLEO}),\ }\href {https://doi.org/10.1103/PhysRevLett.80.3919} {\bibfield  {journal} {\bibinfo  {journal} {Phys. Rev. Lett.}\ }\textbf {\bibinfo {volume} {80}},\ \bibinfo {pages} {3919} (\bibinfo {year} {1998})},\ \Eprint {https://arxiv.org/abs/hep-ex/9711011} {arXiv:hep-ex/9711011} \BibitemShut {NoStop}%
\bibitem [{\citenamefont {Ablikim}\ \emph {et~al.}(2015)\citenamefont {Ablikim} \emph {et~al.}}]{BESIII:2014rqs}%
  \BibitemOpen
  \bibfield  {author} {\bibinfo {author} {\bibfnamefont {M.}~\bibnamefont {Ablikim}} \emph {et~al.} (\bibinfo {collaboration} {BESIII}),\ }\href {https://doi.org/10.1103/PhysRevD.91.031101} {\bibfield  {journal} {\bibinfo  {journal} {Phys. Rev. D}\ }\textbf {\bibinfo {volume} {91}},\ \bibinfo {pages} {031101} (\bibinfo {year} {2015})},\ \Eprint {https://arxiv.org/abs/1412.4566} {arXiv:1412.4566 [hep-ex]} \BibitemShut {NoStop}%
\bibitem [{\citenamefont {Ablikim}\ \emph {et~al.}(2023)\citenamefont {Ablikim} \emph {et~al.}}]{BESIII:2022kbd}%
  \BibitemOpen
  \bibfield  {author} {\bibinfo {author} {\bibfnamefont {M.}~\bibnamefont {Ablikim}} \emph {et~al.} (\bibinfo {collaboration} {BESIII}),\ }\href {https://doi.org/10.1103/PhysRevD.107.032011} {\bibfield  {journal} {\bibinfo  {journal} {Phys. Rev. D}\ }\textbf {\bibinfo {volume} {107}},\ \bibinfo {pages} {032011} (\bibinfo {year} {2023})},\ \Eprint {https://arxiv.org/abs/2212.13361} {arXiv:2212.13361 [hep-ex]} \BibitemShut {NoStop}%
\bibitem [{\citenamefont {Li}\ \emph {et~al.}(2020)\citenamefont {Li}, \citenamefont {L{\"u}}, \citenamefont {Wang}, \citenamefont {Wang},\ and\ \citenamefont {Wei}}]{Li:2020rcg}%
  \BibitemOpen
  \bibfield  {author} {\bibinfo {author} {\bibfnamefont {H.-D.}\ \bibnamefont {Li}}, \bibinfo {author} {\bibfnamefont {C.-D.}\ \bibnamefont {L{\"u}}}, \bibinfo {author} {\bibfnamefont {C.}~\bibnamefont {Wang}}, \bibinfo {author} {\bibfnamefont {Y.-M.}\ \bibnamefont {Wang}},\ and\ \bibinfo {author} {\bibfnamefont {Y.-B.}\ \bibnamefont {Wei}},\ }\href {https://doi.org/10.1007/JHEP04(2020)023} {\bibfield  {journal} {\bibinfo  {journal} {JHEP}\ }\textbf {\bibinfo {volume} {04}},\ \bibinfo {pages} {023}},\ \Eprint {https://arxiv.org/abs/2002.03825} {arXiv:2002.03825 [hep-ph]} \BibitemShut {NoStop}%
\bibitem [{\citenamefont {Pullin}\ and\ \citenamefont {Zwicky}(2021)}]{Pullin:2021ebn}%
  \BibitemOpen
  \bibfield  {author} {\bibinfo {author} {\bibfnamefont {B.}~\bibnamefont {Pullin}}\ and\ \bibinfo {author} {\bibfnamefont {R.}~\bibnamefont {Zwicky}},\ }\href {https://doi.org/10.1007/JHEP09(2021)023} {\bibfield  {journal} {\bibinfo  {journal} {JHEP}\ }\textbf {\bibinfo {volume} {09}},\ \bibinfo {pages} {023}},\ \Eprint {https://arxiv.org/abs/2106.13617} {arXiv:2106.13617 [hep-ph]} \BibitemShut {NoStop}%
\bibitem [{\citenamefont {Amundson}\ \emph {et~al.}(1992)\citenamefont {Amundson}, \citenamefont {Boyd}, \citenamefont {Jenkins}, \citenamefont {Luke}, \citenamefont {Manohar}, \citenamefont {Rosner}, \citenamefont {Savage},\ and\ \citenamefont {Wise}}]{Amundson:1992yp}%
  \BibitemOpen
  \bibfield  {author} {\bibinfo {author} {\bibfnamefont {J.~F.}\ \bibnamefont {Amundson}}, \bibinfo {author} {\bibfnamefont {C.~G.}\ \bibnamefont {Boyd}}, \bibinfo {author} {\bibfnamefont {E.~E.}\ \bibnamefont {Jenkins}}, \bibinfo {author} {\bibfnamefont {M.~E.}\ \bibnamefont {Luke}}, \bibinfo {author} {\bibfnamefont {A.~V.}\ \bibnamefont {Manohar}}, \bibinfo {author} {\bibfnamefont {J.~L.}\ \bibnamefont {Rosner}}, \bibinfo {author} {\bibfnamefont {M.~J.}\ \bibnamefont {Savage}},\ and\ \bibinfo {author} {\bibfnamefont {M.~B.}\ \bibnamefont {Wise}},\ }\href {https://doi.org/10.1016/0370-2693(92)91341-6} {\bibfield  {journal} {\bibinfo  {journal} {Phys. Lett. B}\ }\textbf {\bibinfo {volume} {296}},\ \bibinfo {pages} {415} (\bibinfo {year} {1992})},\ \Eprint {https://arxiv.org/abs/hep-ph/9209241} {arXiv:hep-ph/9209241} \BibitemShut {NoStop}%
\bibitem [{\citenamefont {Tran}\ \emph {et~al.}(2024)\citenamefont {Tran}, \citenamefont {Ivanov}, \citenamefont {Santorelli},\ and\ \citenamefont {Vo}}]{Tran:2023hrn}%
  \BibitemOpen
  \bibfield  {author} {\bibinfo {author} {\bibfnamefont {C.-T.}\ \bibnamefont {Tran}}, \bibinfo {author} {\bibfnamefont {M.~A.}\ \bibnamefont {Ivanov}}, \bibinfo {author} {\bibfnamefont {P.}~\bibnamefont {Santorelli}},\ and\ \bibinfo {author} {\bibfnamefont {Q.-C.}\ \bibnamefont {Vo}},\ }\href {https://doi.org/10.1088/1674-1137/ad102c} {\bibfield  {journal} {\bibinfo  {journal} {Chin. Phys. C}\ }\textbf {\bibinfo {volume} {48}},\ \bibinfo {pages} {023103} (\bibinfo {year} {2024})},\ \Eprint {https://arxiv.org/abs/2311.15248} {arXiv:2311.15248 [hep-ph]} \BibitemShut {NoStop}%
\bibitem [{\citenamefont {Cheung}\ and\ \citenamefont {Hwang}(2014)}]{Cheung:2014cka}%
  \BibitemOpen
  \bibfield  {author} {\bibinfo {author} {\bibfnamefont {C.-Y.}\ \bibnamefont {Cheung}}\ and\ \bibinfo {author} {\bibfnamefont {C.-W.}\ \bibnamefont {Hwang}},\ }\href {https://doi.org/10.1007/JHEP04(2014)177} {\bibfield  {journal} {\bibinfo  {journal} {JHEP}\ }\textbf {\bibinfo {volume} {04}},\ \bibinfo {pages} {177}},\ \Eprint {https://arxiv.org/abs/1401.3917} {arXiv:1401.3917 [hep-ph]} \BibitemShut {NoStop}%
\bibitem [{\citenamefont {Orsland}\ and\ \citenamefont {Hogaasen}(1999)}]{Orsland:1998de}%
  \BibitemOpen
  \bibfield  {author} {\bibinfo {author} {\bibfnamefont {A.~H.}\ \bibnamefont {Orsland}}\ and\ \bibinfo {author} {\bibfnamefont {H.}~\bibnamefont {Hogaasen}},\ }\href {https://doi.org/10.1007/s100529900042} {\bibfield  {journal} {\bibinfo  {journal} {Eur. Phys. J. C}\ }\textbf {\bibinfo {volume} {9}},\ \bibinfo {pages} {503} (\bibinfo {year} {1999})},\ \Eprint {https://arxiv.org/abs/hep-ph/9812347} {arXiv:hep-ph/9812347} \BibitemShut {NoStop}%
\bibitem [{\citenamefont {Jaus}(1996)}]{Jaus:1996np}%
  \BibitemOpen
  \bibfield  {author} {\bibinfo {author} {\bibfnamefont {W.}~\bibnamefont {Jaus}},\ }\href {https://doi.org/10.1103/PhysRevD.53.1349} {\bibfield  {journal} {\bibinfo  {journal} {Phys. Rev. D}\ }\textbf {\bibinfo {volume} {53}},\ \bibinfo {pages} {1349} (\bibinfo {year} {1996})},\ \bibinfo {note} {[Erratum: Phys.Rev.D 54, 5904 (1996)]}\BibitemShut {NoStop}%
\bibitem [{\citenamefont {Goity}\ and\ \citenamefont {Roberts}(2001)}]{Goity:2000dk}%
  \BibitemOpen
  \bibfield  {author} {\bibinfo {author} {\bibfnamefont {J.~L.}\ \bibnamefont {Goity}}\ and\ \bibinfo {author} {\bibfnamefont {W.}~\bibnamefont {Roberts}},\ }\href {https://doi.org/10.1103/PhysRevD.64.094007} {\bibfield  {journal} {\bibinfo  {journal} {Phys. Rev. D}\ }\textbf {\bibinfo {volume} {64}},\ \bibinfo {pages} {094007} (\bibinfo {year} {2001})},\ \Eprint {https://arxiv.org/abs/hep-ph/0012314} {arXiv:hep-ph/0012314} \BibitemShut {NoStop}%
\bibitem [{\citenamefont {Aliev}\ \emph {et~al.}(1994)\citenamefont {Aliev}, \citenamefont {Iltan},\ and\ \citenamefont {Pak}}]{Aliev:1994nq}%
  \BibitemOpen
  \bibfield  {author} {\bibinfo {author} {\bibfnamefont {T.~M.}\ \bibnamefont {Aliev}}, \bibinfo {author} {\bibfnamefont {E.}~\bibnamefont {Iltan}},\ and\ \bibinfo {author} {\bibfnamefont {N.~K.}\ \bibnamefont {Pak}},\ }\href {https://doi.org/10.1016/0370-2693(94)90606-8} {\bibfield  {journal} {\bibinfo  {journal} {Phys. Lett. B}\ }\textbf {\bibinfo {volume} {334}},\ \bibinfo {pages} {169} (\bibinfo {year} {1994})}\BibitemShut {NoStop}%
\bibitem [{\citenamefont {Lu}\ \emph {et~al.}(2024)\citenamefont {Lu}, \citenamefont {Yu}, \citenamefont {Wang},\ and\ \citenamefont {Wu}}]{Lu:2024tgy}%
  \BibitemOpen
  \bibfield  {author} {\bibinfo {author} {\bibfnamefont {J.}~\bibnamefont {Lu}}, \bibinfo {author} {\bibfnamefont {G.-L.}\ \bibnamefont {Yu}}, \bibinfo {author} {\bibfnamefont {Z.-G.}\ \bibnamefont {Wang}},\ and\ \bibinfo {author} {\bibfnamefont {B.}~\bibnamefont {Wu}},\ }\href {https://doi.org/10.1016/j.physletb.2024.138624} {\bibfield  {journal} {\bibinfo  {journal} {Phys. Lett. B}\ }\textbf {\bibinfo {volume} {852}},\ \bibinfo {pages} {138624} (\bibinfo {year} {2024})},\ \Eprint {https://arxiv.org/abs/2401.00669} {arXiv:2401.00669 [hep-ph]} \BibitemShut {NoStop}%
\bibitem [{\citenamefont {Becirevic}\ and\ \citenamefont {Haas}(2011)}]{Becirevic:2009xp}%
  \BibitemOpen
  \bibfield  {author} {\bibinfo {author} {\bibfnamefont {D.}~\bibnamefont {Becirevic}}\ and\ \bibinfo {author} {\bibfnamefont {B.}~\bibnamefont {Haas}},\ }\href {https://doi.org/10.1140/epjc/s10052-011-1734-y} {\bibfield  {journal} {\bibinfo  {journal} {Eur. Phys. J. C}\ }\textbf {\bibinfo {volume} {71}},\ \bibinfo {pages} {1734} (\bibinfo {year} {2011})},\ \Eprint {https://arxiv.org/abs/0903.2407} {arXiv:0903.2407 [hep-lat]} \BibitemShut {NoStop}%
\bibitem [{\citenamefont {Donald}\ \emph {et~al.}(2014)\citenamefont {Donald}, \citenamefont {Davies}, \citenamefont {Koponen},\ and\ \citenamefont {Lepage}}]{Donald:2013sra}%
  \BibitemOpen
  \bibfield  {author} {\bibinfo {author} {\bibfnamefont {G.~C.}\ \bibnamefont {Donald}}, \bibinfo {author} {\bibfnamefont {C.~T.~H.}\ \bibnamefont {Davies}}, \bibinfo {author} {\bibfnamefont {J.}~\bibnamefont {Koponen}},\ and\ \bibinfo {author} {\bibfnamefont {G.~P.}\ \bibnamefont {Lepage}},\ }\href {https://doi.org/10.1103/PhysRevLett.112.212002} {\bibfield  {journal} {\bibinfo  {journal} {Phys. Rev. Lett.}\ }\textbf {\bibinfo {volume} {112}},\ \bibinfo {pages} {212002} (\bibinfo {year} {2014})},\ \Eprint {https://arxiv.org/abs/1312.5264} {arXiv:1312.5264 [hep-lat]} \BibitemShut {NoStop}%
\bibitem [{\citenamefont {Meng}\ \emph {et~al.}(2024)\citenamefont {Meng}, \citenamefont {Dang}, \citenamefont {Liu}, \citenamefont {Liu}, \citenamefont {Shen}, \citenamefont {Yan},\ and\ \citenamefont {Zhang}}]{Meng:2024gpd}%
  \BibitemOpen
  \bibfield  {author} {\bibinfo {author} {\bibfnamefont {Y.}~\bibnamefont {Meng}}, \bibinfo {author} {\bibfnamefont {J.-L.}\ \bibnamefont {Dang}}, \bibinfo {author} {\bibfnamefont {C.}~\bibnamefont {Liu}}, \bibinfo {author} {\bibfnamefont {Z.}~\bibnamefont {Liu}}, \bibinfo {author} {\bibfnamefont {T.}~\bibnamefont {Shen}}, \bibinfo {author} {\bibfnamefont {H.}~\bibnamefont {Yan}},\ and\ \bibinfo {author} {\bibfnamefont {K.-L.}\ \bibnamefont {Zhang}},\ }\href {https://doi.org/10.1103/PhysRevD.109.074511} {\bibfield  {journal} {\bibinfo  {journal} {Phys. Rev. D}\ }\textbf {\bibinfo {volume} {109}},\ \bibinfo {pages} {074511} (\bibinfo {year} {2024})},\ \Eprint {https://arxiv.org/abs/2401.13475} {arXiv:2401.13475 [hep-lat]} \BibitemShut {NoStop}%
\bibitem [{\citenamefont {Frezzotti}\ \emph {et~al.}(2023)\citenamefont {Frezzotti}, \citenamefont {Tantalo}, \citenamefont {Gagliardi}, \citenamefont {Sanfilippo}, \citenamefont {Simula}, \citenamefont {Lubicz}, \citenamefont {Mazzetti}, \citenamefont {Martinelli},\ and\ \citenamefont {Sachrajda}}]{Frezzotti:2023ygt}%
  \BibitemOpen
  \bibfield  {author} {\bibinfo {author} {\bibfnamefont {R.}~\bibnamefont {Frezzotti}}, \bibinfo {author} {\bibfnamefont {N.}~\bibnamefont {Tantalo}}, \bibinfo {author} {\bibfnamefont {G.}~\bibnamefont {Gagliardi}}, \bibinfo {author} {\bibfnamefont {F.}~\bibnamefont {Sanfilippo}}, \bibinfo {author} {\bibfnamefont {S.}~\bibnamefont {Simula}}, \bibinfo {author} {\bibfnamefont {V.}~\bibnamefont {Lubicz}}, \bibinfo {author} {\bibfnamefont {F.}~\bibnamefont {Mazzetti}}, \bibinfo {author} {\bibfnamefont {G.}~\bibnamefont {Martinelli}},\ and\ \bibinfo {author} {\bibfnamefont {C.~T.}\ \bibnamefont {Sachrajda}},\ }\href {https://doi.org/10.1103/PhysRevD.108.074505} {\bibfield  {journal} {\bibinfo  {journal} {Phys. Rev. D}\ }\textbf {\bibinfo {volume} {108}},\ \bibinfo {pages} {074505} (\bibinfo {year} {2023})},\ \Eprint {https://arxiv.org/abs/2306.05904} {arXiv:2306.05904 [hep-lat]} \BibitemShut {NoStop}%
\bibitem [{\citenamefont {Hu}\ \emph {et~al.}(2024)\citenamefont {Hu} \emph {et~al.}}]{CLQCD:2023sdb}%
  \BibitemOpen
  \bibfield  {author} {\bibinfo {author} {\bibfnamefont {Z.-C.}\ \bibnamefont {Hu}} \emph {et~al.} (\bibinfo {collaboration} {CLQCD}),\ }\href {https://doi.org/10.1103/PhysRevD.109.054507} {\bibfield  {journal} {\bibinfo  {journal} {Phys. Rev. D}\ }\textbf {\bibinfo {volume} {109}},\ \bibinfo {pages} {054507} (\bibinfo {year} {2024})},\ \Eprint {https://arxiv.org/abs/2310.00814} {arXiv:2310.00814 [hep-lat]} \BibitemShut {NoStop}%
\bibitem [{\citenamefont {Lees}\ \emph {et~al.}(2013)\citenamefont {Lees} \emph {et~al.}}]{BaBar:2013thi}%
  \BibitemOpen
  \bibfield  {author} {\bibinfo {author} {\bibfnamefont {J.~P.}\ \bibnamefont {Lees}} \emph {et~al.} (\bibinfo {collaboration} {BaBar}),\ }\href {https://doi.org/10.1103/PhysRevLett.111.111801} {\bibfield  {journal} {\bibinfo  {journal} {Phys. Rev. Lett.}\ }\textbf {\bibinfo {volume} {111}},\ \bibinfo {pages} {111801} (\bibinfo {year} {2013})},\ \Eprint {https://arxiv.org/abs/1304.5657} {arXiv:1304.5657 [hep-ex]} \BibitemShut {NoStop}%
\bibitem [{\citenamefont {Bhattacharya}\ \emph {et~al.}(1996)\citenamefont {Bhattacharya}, \citenamefont {Gupta}, \citenamefont {Kilcup},\ and\ \citenamefont {Sharpe}}]{Bhattacharya:1995fz}%
  \BibitemOpen
  \bibfield  {author} {\bibinfo {author} {\bibfnamefont {T.}~\bibnamefont {Bhattacharya}}, \bibinfo {author} {\bibfnamefont {R.}~\bibnamefont {Gupta}}, \bibinfo {author} {\bibfnamefont {G.}~\bibnamefont {Kilcup}},\ and\ \bibinfo {author} {\bibfnamefont {S.~R.}\ \bibnamefont {Sharpe}},\ }\href {https://doi.org/10.1103/PhysRevD.53.6486} {\bibfield  {journal} {\bibinfo  {journal} {Phys. Rev. D}\ }\textbf {\bibinfo {volume} {53}},\ \bibinfo {pages} {6486} (\bibinfo {year} {1996})},\ \Eprint {https://arxiv.org/abs/hep-lat/9512021} {arXiv:hep-lat/9512021} \BibitemShut {NoStop}%
\bibitem [{\citenamefont {Be{\v{c}}irevi{\'c}}\ \emph {et~al.}(2015)\citenamefont {Be{\v{c}}irevi{\'c}}, \citenamefont {Kruse},\ and\ \citenamefont {Sanfilippo}}]{Becirevic:2014rda}%
  \BibitemOpen
  \bibfield  {author} {\bibinfo {author} {\bibfnamefont {D.}~\bibnamefont {Be{\v{c}}irevi{\'c}}}, \bibinfo {author} {\bibfnamefont {M.}~\bibnamefont {Kruse}},\ and\ \bibinfo {author} {\bibfnamefont {F.}~\bibnamefont {Sanfilippo}},\ }\href {https://doi.org/10.1007/JHEP05(2015)014} {\bibfield  {journal} {\bibinfo  {journal} {JHEP}\ }\textbf {\bibinfo {volume} {05}},\ \bibinfo {pages} {014}},\ \Eprint {https://arxiv.org/abs/1411.6426} {arXiv:1411.6426 [hep-lat]} \BibitemShut {NoStop}%
\bibitem [{\citenamefont {Liang}\ \emph {et~al.}(2023)\citenamefont {Liang}, \citenamefont {Alexandru}, \citenamefont {Draper}, \citenamefont {Liu}, \citenamefont {Wang}, \citenamefont {Wang},\ and\ \citenamefont {Yang}}]{Liang:2023jfj}%
  \BibitemOpen
  \bibfield  {author} {\bibinfo {author} {\bibfnamefont {J.}~\bibnamefont {Liang}}, \bibinfo {author} {\bibfnamefont {A.}~\bibnamefont {Alexandru}}, \bibinfo {author} {\bibfnamefont {T.}~\bibnamefont {Draper}}, \bibinfo {author} {\bibfnamefont {K.-F.}\ \bibnamefont {Liu}}, \bibinfo {author} {\bibfnamefont {B.}~\bibnamefont {Wang}}, \bibinfo {author} {\bibfnamefont {G.}~\bibnamefont {Wang}},\ and\ \bibinfo {author} {\bibfnamefont {Y.-B.}\ \bibnamefont {Yang}} (\bibinfo {collaboration} {{\ensuremath{\chi}}QCD}),\ }\href {https://doi.org/10.1103/PhysRevD.108.094512} {\bibfield  {journal} {\bibinfo  {journal} {Phys. Rev. D}\ }\textbf {\bibinfo {volume} {108}},\ \bibinfo {pages} {094512} (\bibinfo {year} {2023})},\ \Eprint {https://arxiv.org/abs/2301.04331} {arXiv:2301.04331 [hep-lat]} \BibitemShut {NoStop}%
\bibitem [{\citenamefont {Hill}\ and\ \citenamefont {Paz}(2010)}]{Hill:2010yb}%
  \BibitemOpen
  \bibfield  {author} {\bibinfo {author} {\bibfnamefont {R.~J.}\ \bibnamefont {Hill}}\ and\ \bibinfo {author} {\bibfnamefont {G.}~\bibnamefont {Paz}},\ }\href {https://doi.org/10.1103/PhysRevD.82.113005} {\bibfield  {journal} {\bibinfo  {journal} {Phys. Rev. D}\ }\textbf {\bibinfo {volume} {82}},\ \bibinfo {pages} {113005} (\bibinfo {year} {2010})},\ \Eprint {https://arxiv.org/abs/1008.4619} {arXiv:1008.4619 [hep-ph]} \BibitemShut {NoStop}%
\bibitem [{\citenamefont {Sint}\ and\ \citenamefont {Weisz}(1997)}]{Sint:1997jx}%
  \BibitemOpen
  \bibfield  {author} {\bibinfo {author} {\bibfnamefont {S.}~\bibnamefont {Sint}}\ and\ \bibinfo {author} {\bibfnamefont {P.}~\bibnamefont {Weisz}},\ }\href {https://doi.org/10.1016/S0550-3213(97)00372-6} {\bibfield  {journal} {\bibinfo  {journal} {Nucl. Phys. B}\ }\textbf {\bibinfo {volume} {502}},\ \bibinfo {pages} {251} (\bibinfo {year} {1997})},\ \Eprint {https://arxiv.org/abs/hep-lat/9704001} {arXiv:hep-lat/9704001} \BibitemShut {NoStop}%
\bibitem [{\citenamefont {Gerardin}\ \emph {et~al.}(2019)\citenamefont {Gerardin}, \citenamefont {Harris},\ and\ \citenamefont {Meyer}}]{Gerardin:2018kpy}%
  \BibitemOpen
  \bibfield  {author} {\bibinfo {author} {\bibfnamefont {A.}~\bibnamefont {Gerardin}}, \bibinfo {author} {\bibfnamefont {T.}~\bibnamefont {Harris}},\ and\ \bibinfo {author} {\bibfnamefont {H.~B.}\ \bibnamefont {Meyer}},\ }\href {https://doi.org/10.1103/PhysRevD.99.014519} {\bibfield  {journal} {\bibinfo  {journal} {Phys. Rev. D}\ }\textbf {\bibinfo {volume} {99}},\ \bibinfo {pages} {014519} (\bibinfo {year} {2019})},\ \Eprint {https://arxiv.org/abs/1811.08209} {arXiv:1811.08209 [hep-lat]} \BibitemShut {NoStop}%
\bibitem [{\citenamefont {Heitger}\ and\ \citenamefont {Joswig}(2021)}]{Heitger:2020zaq}%
  \BibitemOpen
  \bibfield  {author} {\bibinfo {author} {\bibfnamefont {J.}~\bibnamefont {Heitger}}\ and\ \bibinfo {author} {\bibfnamefont {F.}~\bibnamefont {Joswig}} (\bibinfo {collaboration} {ALPHA}),\ }\href {https://doi.org/10.1140/epjc/s10052-021-09037-4} {\bibfield  {journal} {\bibinfo  {journal} {Eur. Phys. J. C}\ }\textbf {\bibinfo {volume} {81}},\ \bibinfo {pages} {254} (\bibinfo {year} {2021})},\ \Eprint {https://arxiv.org/abs/2010.09539} {arXiv:2010.09539 [hep-lat]} \BibitemShut {NoStop}%
\bibitem [{\citenamefont {Abumusabh}\ \emph {et~al.}(2025)\citenamefont {Abumusabh} \emph {et~al.}}]{Belle-II:2025dzk}%
  \BibitemOpen
  \bibfield  {author} {\bibinfo {author} {\bibfnamefont {M.}~\bibnamefont {Abumusabh}} \emph {et~al.} (\bibinfo {collaboration} {Belle-II}),\ }\href@noop {} {\  (\bibinfo {year} {2025})},\ \Eprint {https://arxiv.org/abs/2510.27174} {arXiv:2510.27174 [hep-ex]} \BibitemShut {NoStop}%
\end{thebibliography}%
\onecolumngrid
\clearpage
\begin{center}
    \vspace{1cm}
    \textbf{\large Appendix: Additional figures and tables}
    \vspace{0.5cm}
\end{center}
\begin{table*}[htbp]  
\normalsize 
\centering
 
\renewcommand{\arraystretch}{1.5} 
 
\setlength{\tabcolsep}{12pt}

\begin{tabular}{@{} l ccc @{}} %
\toprule
SVD cut & $V_{D^{\ast+}D^+\gamma}$   & $V_{D_s^{\ast}D_s\gamma}$  & $V_{D^{\ast0}D^0\gamma}$  \\
\midrule
$-10^{-12}$ & $-0.1657(271)$ & $1.8166(582)$ & $-0.1809(138)$ \\
$-10^{-8}$  & $-0.1657(271)$ & $1.8166(582)$ & $-0.1809(138)$ \\
$-10^{-6}$  & $-0.1657(271)$ & $1.8166(582)$ & $-0.1809(138)$ \\
$-10^{-4}$  & $-0.1650(277)$ & $1.8149(567)$ & $-0.1757(119)$ \\
$-10^{-3}$  & $-0.1658(318)$ & $1.7897(606)$ & $-0.1697(139)$ \\
\bottomrule
\end{tabular}
\caption{The fit results for $V_{D^{\ast+}D^+\gamma}$, $V_{D_s^{\ast}D_s\gamma}$ and $V_{D^{\ast0}D^0\gamma}$ on the F32P21 ensemble with different SVD cuts.}
\label{tab:cut}
\end{table*}
\begin{figure*}[htbp] 
    \centering
    \vspace{0.5cm}   
    \includegraphics[width=0.32\textwidth]{reply_figure/F32P23_cll.pdf}
    \includegraphics[width=0.32\textwidth]{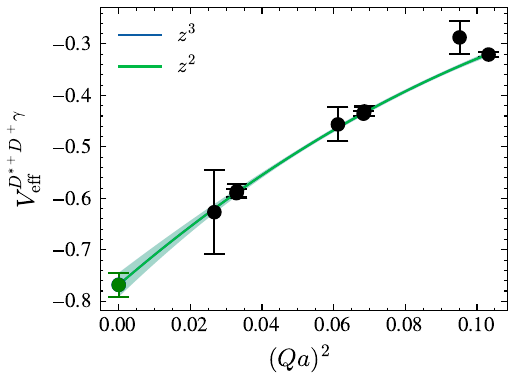}
    \includegraphics[width=0.32\textwidth]{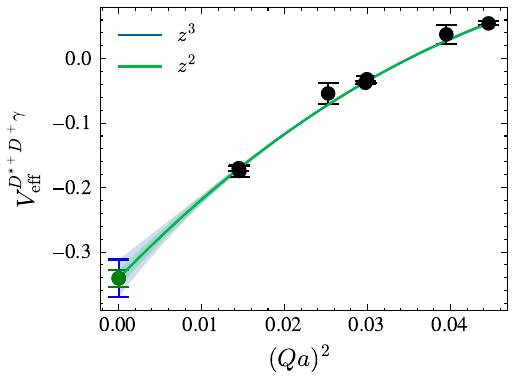}
        
    \includegraphics[width=0.32\textwidth]{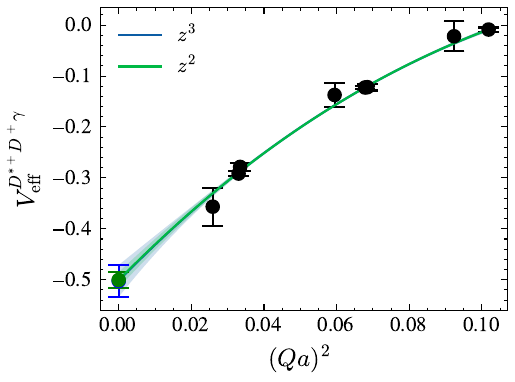}
    \includegraphics[width=0.32\textwidth]{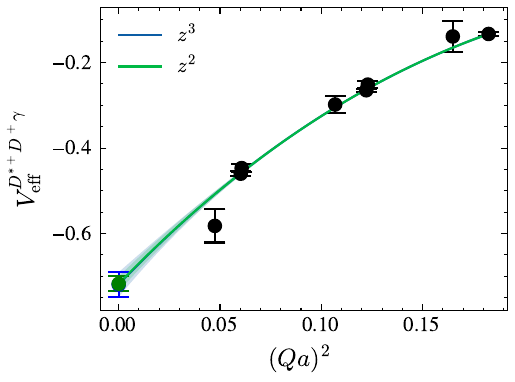}
    \includegraphics[width=0.32\textwidth]{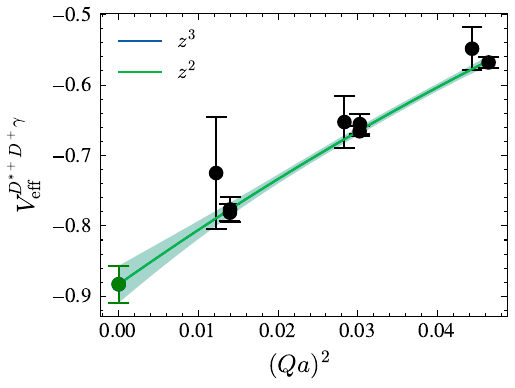}
    
    \caption{Momentum transfer extrapolations for $V^{D^{\ast +} D^+ \gamma}_{\text{eff}}$ for different gauge ensembles. The first row displays the results from the F32P21, C32P23 and H48P32 ensembles from left to right, while the second row is for the F32P30, C24P29 and C48P14 ensembles.}\label{fig:q2e1}
\end{figure*}

\begin{figure*}[h]
    \centering
    \includegraphics[width=0.32\textwidth]{reply_figure/F32P23_clc.pdf}
    \includegraphics[width=0.32\textwidth]{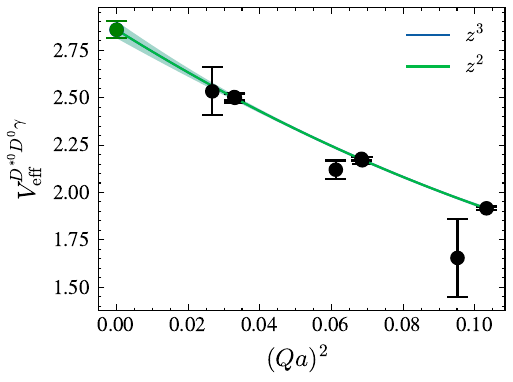}
    \includegraphics[width=0.32\textwidth]{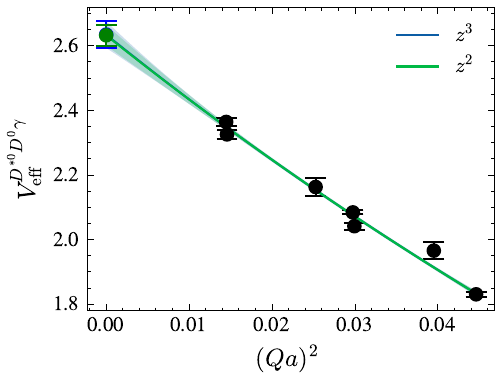}
    
    \includegraphics[width=0.32\textwidth]{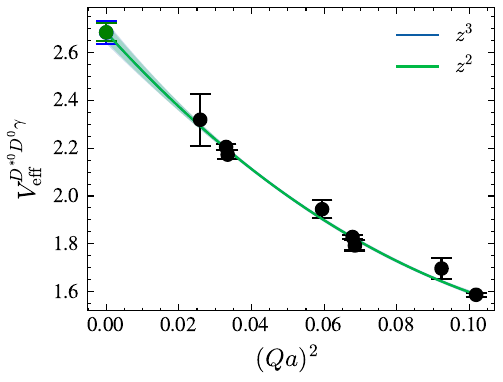}
    \includegraphics[width=0.32\textwidth]{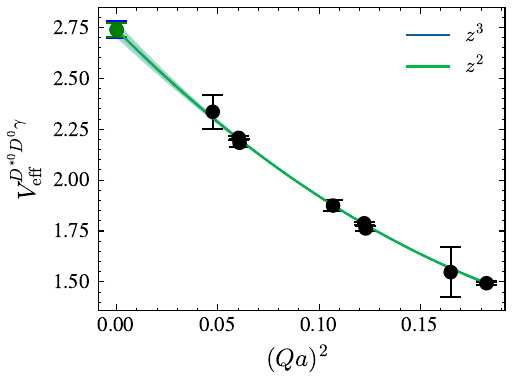}
    \includegraphics[width=0.32\textwidth]{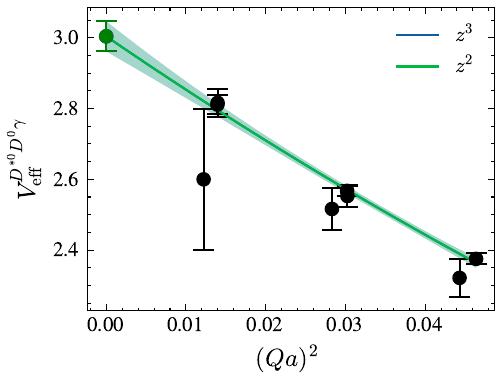}   
    \caption{The same as Fig.\ref{fig:q2e1} but for $V_{\text{eff}}^{D^{\ast 0} D^0 \gamma}$.}
    \label{fig:q2e2}
\end{figure*}

\begin{figure*}[h]
    \centering
    \includegraphics[width=0.32\textwidth]{reply_figure/F32P23_cls.pdf}
    \includegraphics[width=0.32\textwidth]{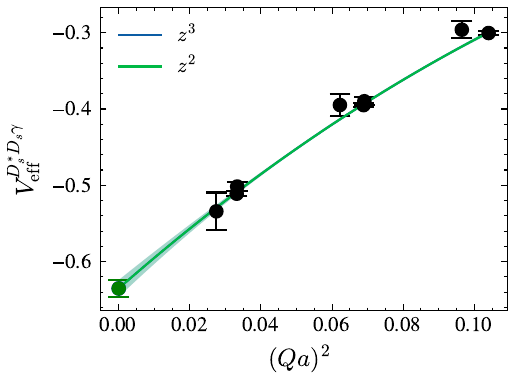}
    \includegraphics[width=0.32\textwidth]{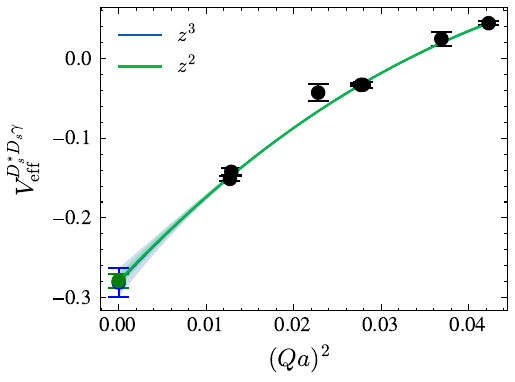}
      
    \includegraphics[width=0.32\textwidth]{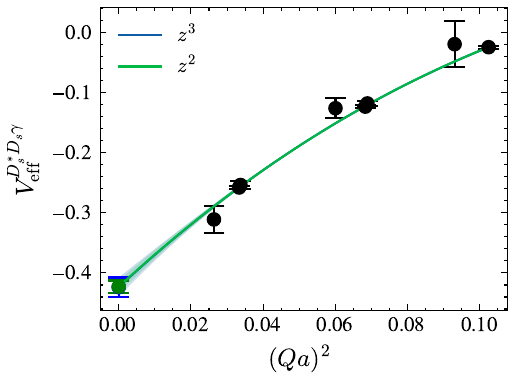}
    \includegraphics[width=0.32\textwidth]{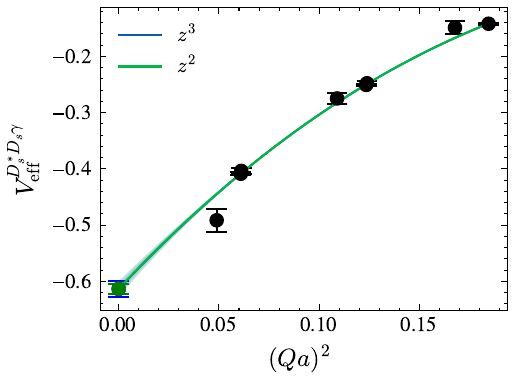}
    \includegraphics[width=0.32\textwidth]{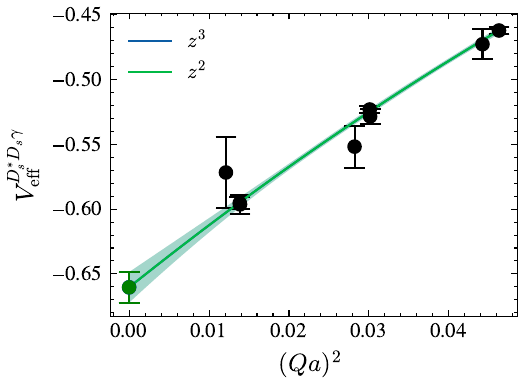}     
    \caption{The same as Fig.\ref{fig:q2e1} but for $V_{\text{eff}}^{D_s^{\ast +} D^+_s \gamma}$.}
    \label{fig:q2e3}
\end{figure*}

\begin{table*}
\centering
\label{tab:combined_data}
\begin{tabular}{cccccc}
\toprule
Ensemble & $(Qa)_D^2$ & $V_{\text{eff}}^{D^{\ast +} D^+ \gamma }$ & $V_{\text{eff}}^{D^{\ast 0} D^0 \gamma }$  & $(Qa)_{D_s}^2$ & $V_{\text{eff}}^{D_s^{\ast +} D^+_s \gamma }$   \\
\midrule 

C48P14 & 0.0122 & -0.7249(799) & 2.5999(2000) & 0.0121 & -0.5718(274) \\
       & 0.0140 & -0.7812(122) & 2.8112(259)  & 0.0138 & -0.5954(49)  \\
       & 0.0140 & -0.7766(180) & 2.8161(399)  & 0.0139 & -0.5964(71)  \\
       & 0.0283 & -0.6526(372) & 2.5162(592)  & 0.0282 & -0.5518(162) \\
       & 0.0302 & -0.6657(69)  & 2.5675(144)  & 0.0301 & -0.5232(25)  \\
       & 0.0303 & -0.6554(138) & 2.5529(310)  & 0.0301 & -0.5285(56)  \\
       & 0.0443 & -0.5484(306) & 2.3215(538)  & 0.0442 & -0.4728(115) \\
       & 0.0464 & -0.5679(74)  & 2.3751(154)  & 0.0463 & -0.4623(27)  \\
\cmidrule{1-6}
C32P23 & 0.0267 & -0.6269(820) & 2.5330(1265) & 0.0275 & -0.5342(248) \\
       & 0.0329 & -0.5899(79)  & 2.5027(163)  & 0.0332 & -0.5111(36)  \\
       & 0.0330 & -0.5863(136) & 2.4990(249)  & 0.0333 & -0.5017(63)  \\
       & 0.0612 & -0.4563(329) & 2.1202(485)  & 0.0622 & -0.3948(148) \\
       & 0.0683 & -0.4355(44)  & 2.1774(89)   & 0.0689 & -0.3949(22)  \\
       & 0.0686 & -0.4312(97)  & 2.1697(186)  & 0.0691 & -0.3898(50)  \\
       & 0.0951 & -0.2874(314) & 1.6551(2054) & 0.0965 & -0.2960(110) \\
       & 0.1032 & -0.3206(51)  & 1.9166(96)   & 0.1040 & -0.3004(25)  \\
\cmidrule{1-6}
C24P29 & 0.0475 & -0.5822(388) & 2.3353(827)  & 0.0489 & -0.4915(204) \\
       & 0.0602 & -0.4600(53)  & 2.2076(108)  & 0.0609 & -0.4087(26)  \\
       & 0.0607 & -0.4469(98)  & 2.1833(190)  & 0.0611 & -0.4037(48)  \\
       & 0.1069 & -0.2982(199) & 1.8751(278)  & 0.1089 & -0.2748(101) \\
       & 0.1222 & -0.2648(32)  & 1.7880(62)   & 0.1234 & -0.2508(16)  \\
       & 0.1230 & -0.2515(80)  & 1.7625(139)  & 0.1238 & -0.2477(40)  \\
       & 0.1649 & -0.1389(355) & 1.5482(1218) & 0.1677 & -0.1489(114) \\
       & 0.1826 & -0.1331(38)  & 1.4935(86)   & 0.1845 & -0.1419(21)  \\
\cmidrule{1-6}
F32P21     & 0.0242 & -0.4971(508) & 2.6672(1113) & 0.0257 & -0.3642(290) \\
           & 0.0326 & -0.3199(56)  & 2.2509(130)  & 0.0333 & -0.2782(30)  \\
           & 0.0329 & -0.3333(106) & 2.2502(213)  & 0.0336 & -0.2751(53)  \\
           & 0.0570 & -0.2282(396) & 1.9664(855)  & 0.0589 & -0.1416(253) \\
           & 0.0670 & -0.1523(34)  & 1.8848(75)   & 0.0681 & -0.1444(18)  \\
           & 0.0675 & -0.1555(81)  & 1.8639(169)  & 0.0685 & -0.1400(43)  \\
           & 0.0890 & -0.1081(250) & 1.6595(1179) & 0.0915 & -0.0896(143) \\
           & 0.1003 & -0.0390(42)  & 1.6159(98)   & 0.1019 & -0.0531(23)  \\
 \cmidrule{1-6}
F32P30 & 0.0259 & -0.3565(371) & 2.3183(1085) & 0.0264 & -0.3109(219) \\
       & 0.0330 & -0.2915(46)  & 2.2046(108)  & 0.0333 & -0.2576(31)  \\
       & 0.0334 & -0.2784(82)  & 2.1727(187)  & 0.0337 & -0.2535(60)  \\
       & 0.0594 & -0.1371(234) & 1.9448(377)  & 0.0600 & -0.1256(165) \\
       & 0.0678 & -0.1225(30)  & 1.8288(85)   & 0.0683 & -0.1230(20)  \\
       & 0.0685 & -0.1219(66)  & 1.7935(210)  & 0.0689 & -0.1177(48)  \\
       & 0.0923 & -0.0219(294) & 1.6973(449)  & 0.0930 & -0.0190(381) \\
       & 0.1018 & -0.0087(40)  & 1.5876(88)   & 0.1024 & -0.0240(25)  \\

\cmidrule{1-6}
H48P32 & 0.0145 & -0.1699(41)  & 2.3643(114)  & 0.0127 & -0.1511(31)  \\
       & 0.0146 & -0.1743(86)  & 2.3256(136)  & 0.0129 & -0.1422(46)  \\
       & 0.0252 & -0.0540(158) & 2.1630(290)  & 0.0228 & -0.0428(106) \\
       & 0.0297 & -0.0375(24)  & 2.0843(63)   & 0.0277 & -0.0332(18)  \\
       & 0.0299 & -0.0320(53)  & 2.0417(107)  & 0.0279 & -0.0330(37)  \\
       & 0.0395 &  0.0376(142) & 1.9660(261)  & 0.0369 &  0.0245(89)  \\
       & 0.0446 &  0.0547(31)  & 1.8308(80)   & 0.0423 &  0.0442(23)  \\
\bottomrule
\end{tabular}
\caption{The values of $(Qa)^2$ and the effective form factors fitted on each ensemble. $(Qa)_D$ denotes the momentum transfers for the $D^{\ast +} \rightarrow D^+\gamma$ and $D^{\ast 0} \rightarrow D^0\gamma$ processes. While, $(Qa)_{D_s}$ denotes the momentum transfer for the $D_s^{\ast +} \rightarrow D^+_s\gamma$ process.}\label{tab:fits}
\end{table*}

\clearpage

\begin{figure}[htbp]
    \centering
    \includegraphics[width=0.45\textwidth]{reply_figure/cl_l_pi_noa.pdf}
    \includegraphics[width=0.45\textwidth]{reply_figure/cl_l_pi_ln.pdf}
    \includegraphics[width=0.45\textwidth]{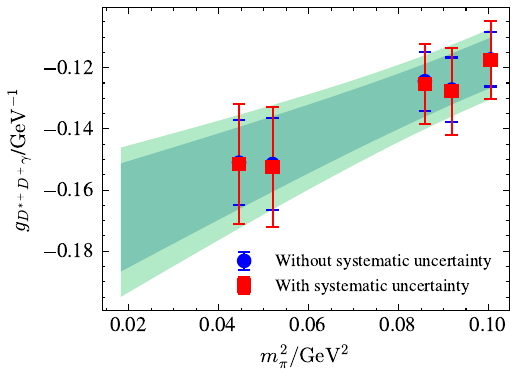}
    \includegraphics[width=0.45\textwidth]{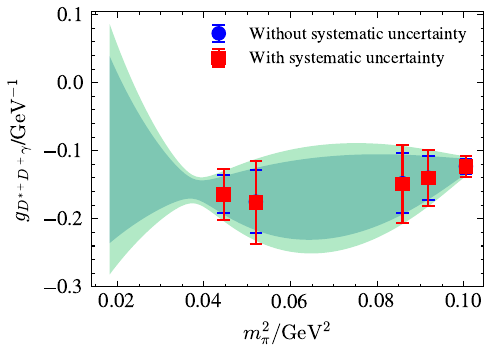}    
    \caption{The upper panels present the results including the C48P14 ensemble, while the lower panels show the results excluding it. The left and right columns correspond to extrapolations with Eq.~(\ref{eq19}) and Eq.~(\ref{eq20}) respectively.}
  
    \label{fig:coupling_compare}
\end{figure}

\end{document}